\documentstyle[psfig]{l-aa}
\begin{document}  
\thesaurus{
           }

\title{HST images and properties of the most distant radio galaxies}

\author{L. Pentericci\inst{1}\and
        H.J.A. R\"ottgering\inst{1}\and
        G.K. Miley\inst{1}\and
        P. McCarthy\inst{2}\and
        H. Spinrad\inst{3}\and
        W.J.M van Breugel\inst{4}\and
        F. Macchetto\inst{5}}

\offprints{L. Pentericci}

\institute{Leiden Observatory, P.O.~Box 9513, NL - 2300 RA Leiden,
The Netherlands \and
The Observatories of the Carnegie Institution of Washington,813 Santa Barbara Street, Pasadena, California 91101 \and
Astronomy Department, University of California, Berkeley, CA 94720 \and
Lawrence Livermore Laboratory, PO Box 808, Livermore, CA 94459 \and
Space Telescope Science Institute, 3700 San Martin Drive, Baltimore, MD 21218
}

\date{Version 6 July}

\maketitle

\markboth{L. Pentericci et al.: HST images and properties of HZRGs.}{} 

\begin{abstract}
With the Hubble Space Telescope we have obtained images of 9 of the most distant radio galaxies. The galaxies, which have redshifts between 
$z=2.3$ and $z=3.6$, were observed with the  
WFPC2 camera in a broad band filter 
(F606W or F707W, roughly equivalent to V or R-band),
corresponding to the near ultraviolet emission in the rest frame 
of the radio galaxies. The total observing time was 2 orbits per object. 
In this paper we present the images overlayed on 
VLA radio maps of comparable resolution. We also present previously 
unpublished images, taken from the HST archive, of two other high 
redshift radio galaxies,
observed through similar broad band filters.
We find that on the scale of the HST observations 
there is a wide variety of morphological structures of the hosting galaxies: 
most objects have a clumpy, irregular appearance, consisting 
of a bright nucleus  and a number of smaller components, suggestive
of merging systems.   
Some observed  structures could be due (at least partly) to the 
presence of dust
distributed through the galaxies. 
The UV continuum emission is generally elongated and aligned with 
the axis of the radio sources, however the characteristics of the 
``alignment effect'' differ from case to case, suggesting that the phenomenon 
cannot be explained by a single physical mechanism. 
We compare the properties of  our radio galaxies with those of the 
 UV dropout galaxies and conclude that (i) the most massive radio galaxies may well evolve from an aggregate of UV dropout galaxies and (ii) high redshift 
radio galaxies probably evolve into present day brightest  cluster galaxies. 
\end{abstract} 
\section{\bf Introduction}
Studying the optical morphology of high redshift ($z > 2$) 
radio galaxies (HZRGs) 
can contribute substantially to our understanding of galaxy formation 
and evolution in the early universe (for a recent review see McCarthy 1993). 
Although the recent development of new techniques (e.g. U and B band dropouts, 
Steidel et al. 1996) has led to the discovery 
of a large population  of high redshift 
galaxies, radio galaxies remain  still  of exceptional 
interest, because they pinpoint the most massive systems at 
high redshift and are potential signposts for finding high-redshift 
clusters of galaxies.
\\
It has been shown that high luminosity radio sources associated
with quasars and radio galaxies at redshift $\sim 0.5$ are located
in rich clusters (e.g. Hill \& Lilly 1991).  
\nocite{hil91} 
At $z \sim 1$ there are now several possible X-ray clusters 
that have been discovered around powerful radio galaxies,
such as 3C324 at z$=$1.2 (Dickinson et al. 1998),
3C356 and 3C280 (Crawford \& Fabian 1996).\nocite{cra96}
At $z>2$ the existence of clusters around HZRGs has not been established.
However, there is an increasing number of important observational 
indications that HZRGs might be in clusters,
including (i) the detection of possibly extended X-ray emission from
the radio galaxy PKS 1138-262 at z=2.156, most probably coming 
from a hot cluster atmosphere (Carilli et al. 1998); \nocite{car98} 
(ii) strong Faraday polarization and
rotation of the radio emission of some HZRGs which might be due to dense 
gaseous halos (Carilli et al. 1997)\nocite{car97};
(iii) possible excess of companion galaxies detected along the axes of the
radio sources (R\"ottgering  et al. 1996)\nocite{rot96e}; (iv) possible 
excess
of Lyman break selected galaxies in the fields 
of several powerful radio sources 
(e.g. Lacy \& Rawlings, 1996) \nocite{lac96} and (v) 
excess of candidate companion galaxies 
(with two objects spectroscopically confirmed) 
in the vicinity of MRC 0316-257, at z=3.14 (Le Fevre et al. 1996).
\\
The hosts  of powerful low redshift radio sources have long been identified
with giant elliptical galaxies, containing old stellar population.
The surprising continuity of the K--z  relation between the high redshift 
radio galaxies and the low redshift brightest cluster galaxies 
which shows little scatter up to redshift  of $\sim 4$ (although 
the scatter increases beyond redshift 2, e.g. Eales 
et al. 1997), might indicate that the hosts of powerful 
radio sources  are the most massive galaxies   
know at high--redshifts. Moreover, since HZRGs are probably located in 
forming clusters of galaxies, they could  be the ancestors of 
brightest cluster galaxies. \nocite{eal97}
\\
We have previously presented and discussed HST images of the radio galaxy
PKS 1138-262 at z=2.156, which shows the clumpiest optical morphology 
of all the HZRGs 
imaged  with the HST (Pentericci et al. 1998)\nocite{pen98}.
Our conclusion was that PKS 1138-262 is giant elliptical galaxy at the center of a 
protocluster in the late stages of its  formation.
\\
In this paper  we present HST--WFPC2 images for 9 powerful 
radio galaxies having  redshifts 
 between z$=$2.3 and z$=$3.6.
We also present deep HST archive images of 2 HZRGs observed with 
WFPC2.
We compare the HST images  with  VLA maps of the associated  radio sources 
having similar resolution. 
After discussing  the sample selection (Sect. 2), we  describe  the 
HST imaging and  reduction procedures (Sect. 3), the radio imaging and 
the problem of the relative astrometry between the radio and HST data 
(Sect. 4). In Sect. 5  
we briefly discuss the most important characteristics of each object, also
referring to  previous results  that are  relevant to the interpretation of the new data. 
Finally in Sect. 6 we discuss  some statistical trends of the properties of
these high redshift radio sources, giving a qualitative interpretation. 
We then summarize our main  results and present
our conclusions.
We also include the Appendix new radio images of the radio galaxies 
TX 1707+105 and MRC 2104-242.
\\
Throughout this paper we assume a Hubble constant of H$_0 =50$ km s$^{-1}$ Mpc$^{-1}$ and a deceleration parameter of q$_0 =0.5$. 
\section{\bf Sample selection}
The radio galaxies  were initially  selected from the more than 
60 HZRGs which were known at the commencement of the project (1995) (e.g. van Ojik 1995 and references
therein).  \nocite{oji95} 
Most of these distant radio galaxies were found by observing ultra steep
spectrum radio sources (USS) ($\alpha <-1.0 $, where $\alpha$ is the radio spectral 
index) (van Ojik 1995). 
\\
Objects were selected according to  the following criteria:
(i) bright in the the R band 
(R$<$ 24, i.e. sufficient to be mappable in a reasonable time with the HST);
(ii) amongst the brightest line emitters  (Ly$\alpha$ flux $ >
 $ 10$^{- 15}$ erg s$^{-1}$cm$^{-2}$).
Because of its high redshift, we also included the radio  galaxy MG 2141+192 
(z=3.594) in the sample.
\\
Finally we obtained unpublished HST/WFPC2 images of the radio sources 
B2 0902+343,
at z=3.395, and TX 0828+192, at z=2.572, from the HST archive. 
\\
For a statistical study of the properties of HZRGs it is important to 
enlarge the sample of objects with HST images: we therefore included 
in our analysis the other radio galaxies that
have been imaged  with the HST. 
These include the radio galaxy 4C 41.17 at z=3.8, one of the the 
best studied HZRGs (van Breugel et al. 1998); the radio galaxy  
PKS 1138-262, at z=2.156  that was studied by our group 
(Pentericci et al. 1997, 1998);
\nocite{pen97} MRC 0406-242 at z=2.44 that was 
object of a multi-frequency study, including WFPC imaging in different color 
bands, by Rush et al. (1997)\nocite{rus97}; and 4C 23.58 at z=2.95 
(Chambers et al. 1996a and 1996b). \nocite{cha96a,cha96b} The first two objects were
imaged with the WFPC2 camera, while the last two were imaged with the
pre-refurbishment HST/WFPC. 
Details of the observations can be found in the mentioned papers.
In this way the final sample available for the statistical analysis
of the properties of HZRGs consists of 15 galaxies.
By including also radio galaxies that have been imaged with the 
pre-refurbishment HST and/or for which the total integration times
are considerably different (e.g. the observations of 4C 41.17 are much deeper
than for the other objects), the quality of the images varies within the sample. However given the relatively 
small number of radio galaxies observed, it is  important 
to increase the statistics.
\\
In addition to the HST images, all the radio galaxies in the final sample
have been imaged with the VLA at several frequencies, to study their 
radio-polarimetric properties (\cite{car97}) 
and have Ly$\alpha$ profiles taken with  resolution of $ <$ 100 km s$^{-1}$, 
thus allowing  a detailed study of 
the morphology and kinematics of the ionized gas (\cite{oji97a}). 
For some of the radio 
galaxies, ground-based  narrow band images of the Ly$\alpha$ emission gas, 
and broad band images in various color bands (mostly R-band and K-band)
are also available (see references for individual objects in Sect. 5).
\section{\bf HST imaging}
\subsection{\bf Observations }
Table 1 summarizes the observations.
9 radio galaxies were imaged with the Planetary Camera (PC) of
WFPC2  during Cycle 5 and/or Cycle 6. The PC utilizes an 800 $\times$ 800 pixel Loral CCD as detector with pixel size of $0.0455''$  (\cite{bur95}).
The typical exposure time was 5300 sec (2 orbits) for  each galaxy.
The total observing time was split between two exposures to 
facilitate removal of cosmic ray events.  
\\
The filters used for the observations were chosen to 
avoid contamination from the strong Ly$\alpha$ emission line at 1216\AA \space
and to have the rest frame wavelengths sampled as similar as possible 
throughout the sample.
For the radio galaxies at redshift $z > 2.9$ the filter used was the 
broad-band F707W filter (centered at $\lambda_0=6868$ \AA \space and with a FWHM of  
$\Delta\lambda =1382$ \AA), similar to the Cousins R band ;
for the lower redshift galaxies we used  
the broad-band F606W filter 
($\lambda_0=5934$ \AA \space and $\Delta\lambda =1498$ \AA) which is similar
to the V band. 
\\  
The radio galaxy TX 0828+193 was observed during Cycle 4 with the WFPC2 by Chambers et al., using 
the filter F675W which is centered at $\lambda_0 =  6756$ \AA \space and has a FWHM of $\Delta\lambda = 865$ \AA. The observations were done in polarimetric  mode.
The galaxy was observed using  the WF3 section   
of WFPC2, which utilizes an 800 $\times$ 800 pixel Loral CCD as detector with pixel size of $0.1''$  (\cite{bur95}).
The total exposure time of 10000 s was split 
between ten observations.  
\\
The radio galaxy B2 0902+343  was observed during Cycle 4 by Eisenhardt and Dickinson, using  the PC of WFPC2 with the filter F622W, which is centered at 
$\lambda_0= 6189.9$ \AA \space and has a FWHM of $\Delta\lambda =916$\AA. 
The total exposure time of 21600 s was split 
between nine  observations.  
\\
In the  redshift range observed the continuum emission may include contribution
from the faint emission lines  of HeII, CIII] and CIV.
For most of the radio galaxies we could estimate  the 
total  contamination using the line fluxes measured by 
low resolution spectra of the objects taken by van Ojik (1995) 
\nocite{oji95}.
The detected lines are listed in Table 1, as well as the total contribution
of the line emission to the measured flux, which ranges from 0 to 13.7\%, 
with the  highest contribution for the radio galaxy TX 0211-122. 
We expect that for the 4 sources of which we do 
not have any such data available, the line contribution will be in the same 
range.    
Therefore we can assume that the images represent to a  good approximation
 the continuum emission from the galaxies.
\begin{table*}
\caption{\large {\bf Observation log}}\label{tab.pol}
\begin{center}
\begin{tabular}{lcccccccccc}
\hline 
Cat.& Source  & z  & Obs. & Filter & Rest.$\lambda\lambda$ & Exp.  & N & WFPC & Lines & \%flux$^{\it a}$ 
\\
    &         &    & date &        &   & time  &   & mag  &  
\\
    &         &    &      &        &  \AA  & sec.  
\\
(1) &  (2)    &(3) & (4)  & (5)    &(6)     &(7)    &(8)&  (9)      &(10) & (11)  
\\
TX & 0211$-$122&2.336&14/8/95& F606W &1554-2003&5300 &2 &22.9& HeII,CIII]  & 13.7  
\\ 
TX & 1707+105& 2.349&7/8/95 & F606W &1550-1998&5300 &2 &23.7$^{\it b}$&HeII,CIII] & 4.2$^{\it c}$
\\
4C & 1410$-$001&2.363&15/8/95& F606W &1542-1987&5300 &2 &22.9& HeII,CIII]  &13.3
\\
MRC & 2104$-$242& 2.491&10/5/97& F606W &1485-1914&5300 &2 &22.5& ---         
& ---
\\
TX  &0828$+$193&2.572 &26/1/96& F675W &1770-2012&10000&10&22.2&  CIII]      & 5.1
\\
MRC &2025$-$218& 2.630 &9/11/95& F606W &1428-1841&5300 &2 &22.6& ---        & 
---
\\ 
4C  &1345+245& 2.879 &29/3/97& F702W &1592-1949&5200 &2 &23.4&  none       & $\le 1$
\\
MRC &0943$-$242 &2.923 &18/11/95& F702W&1575-1927&5300 &2 &22.6&  HeII,CIII] &11.0 
\\
B2 &0902+343&3.395 &5/11/94 & F622W&1304-1512&21600&9 &23.8&  ---        & ---
\\
4C &1243+036& 3.570 &5/8/95  & F702W&1352-1654&5300 &2 &23.2&   none      & $\le 1$ 
\\
MG &2141+192 &3.594 &10/5/97 & F702W&1345-1645&5200 &2 &24.2&  ---        & ---
\\
\hline
\end{tabular}
\end{center}
(1) Catalog. (2) Source name. (3) Redshift. 
(4) Date of HST observations. (5) Filter used for HST observations. (6) 
Rest-frame wavelength interval in angstroms. (7) Total exposure time in second. 
(8) Number of frames in which the total observing time was split. (9) WFPC2 
magnitude within  an aperture of radius 4$''$. (10) Emission line detected within the filter band. (11) Percentual flux contamination from the detected lines. \\ 
$^{\it a}$ See text for a detailed explanation \\ 
$^{\it b}$ Magnitude of galaxy 1707+105A\\
$^{\it c}$ Contribution to the total flux of 1707+105A and 1707+105B.\\
\end{table*}
\subsection{\bf Data Processing}
The data were reduced according to the standard Space Telescope Science Institute pipeline (\cite{lau89}). Further processing was performed using the 
NOAO Image Reduction and Analysis Facility (IRAF) software package 
and the Space Telescope  Science Data Analysis System (STSDAS) and involved  
cosmic ray removal  and registering of the images. The shifts were 
measured from the peak positions of a non-saturated star present 
in both the PC images. The different frames  were then added, 
background subtraction was performed 
using the average 
flux contained within 
4 or more apertures placed on blank areas of the sky, as close as possible
to the source, at different positions, to avoid introducing errors
from residual gradients in the background flux.
The resulting image was 
flux calibrated  according to the precepts described in the 
``HST Data Handbook'' (1995 edition), using the photometric parameters 
from the standard HST calibration and included in the file header.
The images were then rotated to superimpose them to the VLA radio maps 
(see Sect. 4.1)
\\
\begin{table*}
\caption{\large{\bf Properties of the radio sources}}\label{tab.pol}
\hskip1cm
\begin{tabular}{lllllllllll}
\hline
Source &  RA & Dec & F$_{4.5}$& $\alpha$&CF$_{4.5}$&Size& RM & PA &  Ref. 
\\
     &J2000&J2000& mJy     &          & \%       &kpc &rad m$^{-2}$& 
\\
(1) & (2) & (3) & (4) & (5) & (6) & (7) & (8) & (9) & (10) 
\\
\hline
TX 0211$-$122&02:14:17.37&$-$11:58:46.7&54 &1.5&3.8&134& 160&73$^{\it a}$&A 
\\
TX 1707$+$105&17:10:06.85&+10:31:09.0&64 &1.2&-- &173&-- & -58&   C
\\ 
4C 1410$-$001&14:13:15.13&$-$00:22:59.6&57 &1.3&6.7&189&1510&44$^{\it a}$& A
\\ 
MRC 2104$-$242&21:06:58.16&$-$24:05:11.3&68 &1.3&1.6&177&-- & 12&  B
\\
TX 0828$+$193&08:30:53.71&+19:13:18.5&22 &1.6&21 & 98&-- &-42&   A    
\\
MRC 2025$-$218&20:27:59.45&$-$21:40:57.1&95 &1.1&0.7& 38&910& -16& A
\\
4C 1345$+$245&13:48:14.78&+24:15:50.0&115&1.4&0.7& 17&750&-36&  A
\\
MRC 0943$-$242&09:45:32.79&$-$24:28:49.8&55 &1.8&-- & 29&--&75& A
\\
B2 0902$+$343 &09:05:30.10&+34:07:56.9&100&1.4&15 & 32&2500&37 & E
\\   
4C 1243$+$036&12:45:38.43&+03:23:20.3&70 &1.4&2.0& 50&420& 20&  D
\\
MG 2141$+$192&21:44:07.50&+19:29:15.0&67 &1.6&-- & 60&-- & 3  & A
\\
\\
\hline
\end{tabular}
\\
(1) Name of the source (2) and (3) coordinates of the radio core in the epoch J2000. (4) Total radio flux at 4.5 GHz. (5) Radio spectral index between 4.5 GHz and 8.2 GHz, $S_{\nu}= S_0 \nu^{- \alpha}$. (6) Radio core flux /total flux at 4.5 GHz rest-frame. (7) Maximum radio source size. (8) Maximum rotation measure. (9)  Position angle of the 
inner radio axis  relative to the direction North-South. (10) References: 
A. Carilli et al. 1997  B. McCarthy et al. 1990
C. This paper D. van Ojik et al. 1996 E. Carilli 1995
\\
$^{\it a}$ Radio jets might be precessing.\\
\end{table*}
The magnitudes were computed from the unrotated images
(which have less 
smoothing) within a fixed aperture of diameter $4''$.
In most cases this aperture is large enough to enclose all the light from the galaxies.  
The magnitudes were computed as: $m= -2.5 log_{10} F + M(0) $, 
where F is the measured flux and $M(0)=21.1$ is the zero point 
for the HST magnitude scale normalized to Vega. 
The results are presented in Table 1. 
\\
A number of different effects  contributes to the errors in the photometric magnitudes; (i) the Poisson noise of the detected counts; 
(ii) a $\sim$2\% uncertainty 
in  the determination of the zero point (Burrows 1995); 
(iii) a $\sim$4\% systematic error due to the problem of charge transfer efficiency 
in the Loral CCD 
(Holtzman et al. 1995) for which we did not correct; (iv) accurate subtraction of the mean sky background; (v) sky noise within the source aperture.
The last two are usually the predominant effects. 
We estimate that the total  uncertainty
in the magnitudes is  0.1 or less for all galaxies.
\\
A first order transformation from the F606W  and F702W ST magnitudes 
to the standard magnitude system was derived applying 
the precepts described by Holtzman et al. (1995).\nocite{hol95} 
The resulting transformation are $m_V=m_{F602W}+0.25(V-I)$, $m_R=m_{F622}-0.25(V-R)$, $m_V=m_{F675W}+0.21(V-R)$,and $m_R=m_{F702}+0.3(V-R)$.
\\
In Table 1 we list for each galaxy the WFPC magnitude $m$;
the  emission lines that have been detected within the filter band and  
the total line contribution to the continuum flux.  
\section{\bf Radio imaging}
All the radio galaxies  with the exception of B2 0902+343, TX 1707+105 and 
MRC 2104-242
were imaged with the VLA  as part of a high resolution, 
multi-frequency radio polarimetric study 
carried out on a large sample of HZRGs by Carilli et al. (1997).\nocite{car97}
A full description of the observations and the 
reduction procedure can be found in this paper.
\\
The radio map of B2 0902+343 that we use in this paper 
is a high resolution (0.15$''$) radio continuum image of total intensity 
at 1.65GHz obtained by Carilli
by combining data from the VLA and MERLIN (see Carilli 1995 for details)\nocite{car95c}. 
\\
The radio observations of TX 1707+105 and MRC 2104$-$214 were performed 
with the VLA in B array.
Details of observations 
and reduction  for both sources can be found in Appendix A and B.
\subsection{\bf Relative astrometry}
The coordinate frame for the WFPC2 images determined from the 
image header information has uncertainties of the order of $1''$ 
(\cite{bur95}). 
Since the optical galaxies are generally clumpy on a scale smaller 
than $1''$, it is important to get the better possible 
registration between the radio and the optical images,
to allow a detailed inter-comparison between the emissions.
\\
\begin{figure*}
\centerline{
\psfig{file=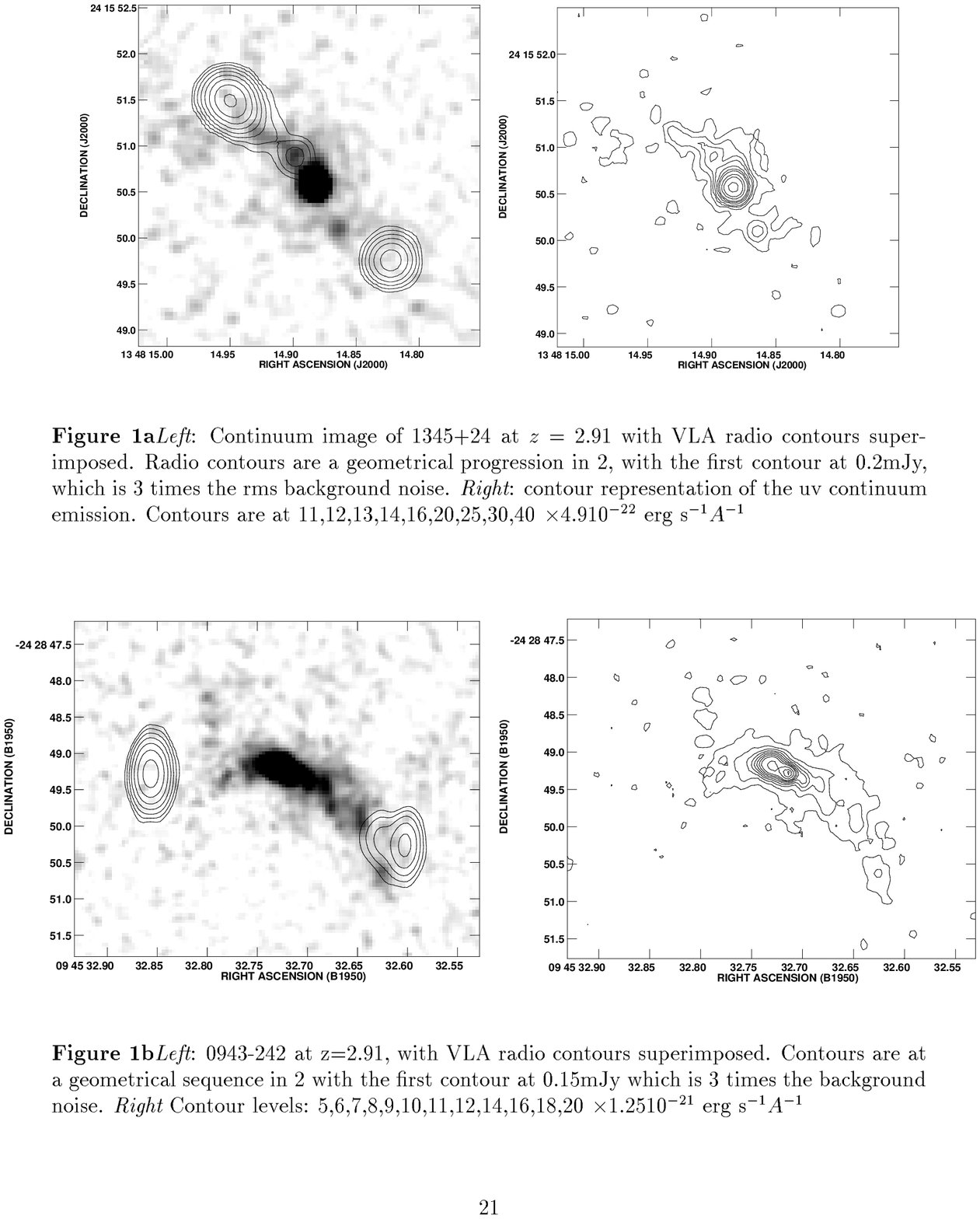,width=7cm,clip=}
\psfig{file=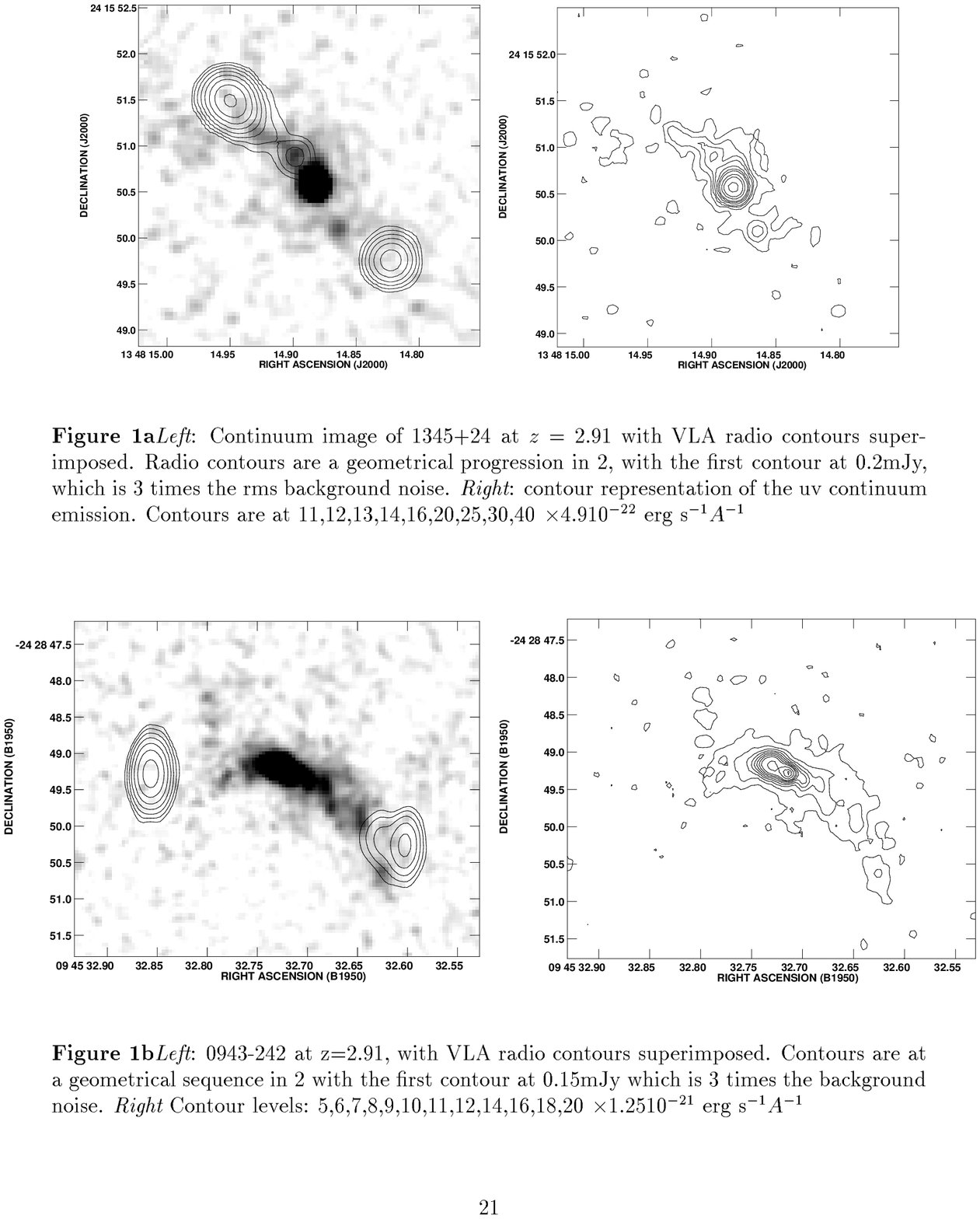,width=7cm,clip=}
}
\caption{{\it Left}: A grey scale continuum image of 4C 1345+245
 at $z=2.879$ with
 VLA radio contours superimposed. Radio contours are a geometrical progression
in 2, with the first contour at 0.2mJy, which is 3 times the rms background noise. {\it Right}: Contour representation of the UV continuum emission.
Contours are at 11,12,13,14,16,20,25,30,40 $\times 4.9 10^{-22}$ erg s$^{-1}$ 
\AA$^{-1}$.}
\end{figure*}
In overlaying the HST images with the radio VLA images
we made the following assumptions: 
for those sources showing a clear detection of the radio nucleus and 
for which  good K-band (or K$_{sh}$-band) images existed (McCarthy, 
private communication), we  
assumed that the peak position of the infra-red image  would  
be a better indicator of the true location of the center of the host
galaxy, rather than the peak of the 
HST image, since the UV continuum might be effected by dust extinction
(e.g de Koff et al. 1996). \nocite{dek96}
We therefore identified  the position of the radio core in the VLA image
with  the peak position of the K-band image. 
Finally we registered the HST frame  and the infrared
frame using the weighted positions of several stars which were present 
on both fields; this can be achieved with an accuracy of 0.1$''$ 
which is then the total final uncertainty in the relative astrometry.
This procedure was possible for the radio galaxies 
TX 0211-122, 4C 1243+036, MRC 2025$-$218 and MRC 2104$-$242.
\\
For those objects which had a clearly detected radio core but no K-band images,
we associated the peak position of the HST image to the peak position
of the radio emission. We followed this procedure for the radio galaxies 4C 1345+245, 4C 1410-001 and TX 0828+128 (for this last object see remarks made in the individual source description): these objects have a relatively simple 
morphology, hence it is reasonable to assume that 
the peak of the UV continuum represents the true nucleus of the galaxy; the final uncertainty of the relative 
astrometry is then within a pixel i.e. $\sim 0.05''$.
\\
For those objects which have no detected radio core 
(MRC 0943$-$242, TX 1707+105
and MG 2141+192) we used the HST absolute astrometry, and we then checked the 
peak position of several stars which were present on the WFPC2 frames, with
the position given in the APM catalog; with this method we achieved 
an accuracy of $\sim 0.8''$. Finally for B2 0902+343 which has a radio 
core but no clear optical nucleus, we kept the natural HST astrometry: in this 
way the radio core falls in between the two optical peaks. This is 
consistent with what found e.g. by Carilli (1995). 
\section{\bf Individual source description}
Grey scale HST WFPC2 images (smoothed with a Gaussian function of FWHM equal 
to 2 pixels) with VLA radio contours superimposed are shown  
in Figs. 1-10. For every source  we also show a contour map 
of the continuum emission to better delineate the morphology.
We do not show such maps for B2 0902+343 and MG 2141+192 
because they have very 
low surface brightness, and a contour map would add no information.
For the very large radio galaxies (namely TX 0211-122, TX 1707+105 and 4C 1410-001)
we also present a third image showing the complete field of the radio source.
The objects are presented in order of increasing radio size, since it has been shown (e.g van Ojik 1995) that  
several properties of HZRGs tend to change with increasing radio size.
\\
We shall now give  brief descriptions of the ultraviolet  
morphology of each radio galaxy,
with special emphasis on any peculiar characteristics (such as distortions, 
jet-like features etc), and compare those with relevant  previous results.
\\
\\
{\bf 4C 1345+245 } 
\\
This radio source at z=2.879 (\cite{cha96b}), is the smallest 
in the sample, being only  $2''$ in extent (corresponding to 17 kpc in 
the adopted cosmology).
The radio structure has been extensively studied with the VLA 
at several frequencies by Carilli et al. (1997), who classified it as  
``compact steep spectrum source'' (CSS) and by Chambers et al. (1996b).\nocite{cha96b} 
The radio emission shows two lobes of roughly equivalent brightness, 
with a one sided feature extending from the core towards the 
eastern side, which has been identified as a jet. 
Optical and infrared ground--based observations show a compact object, 
with the emission extended along the radio axis, with one faint component or companion object along the radio axis to the southwest but beyond the radio 
lobe.  
(\cite{cha96b}).
\begin{figure*}
\centerline{
\psfig{file=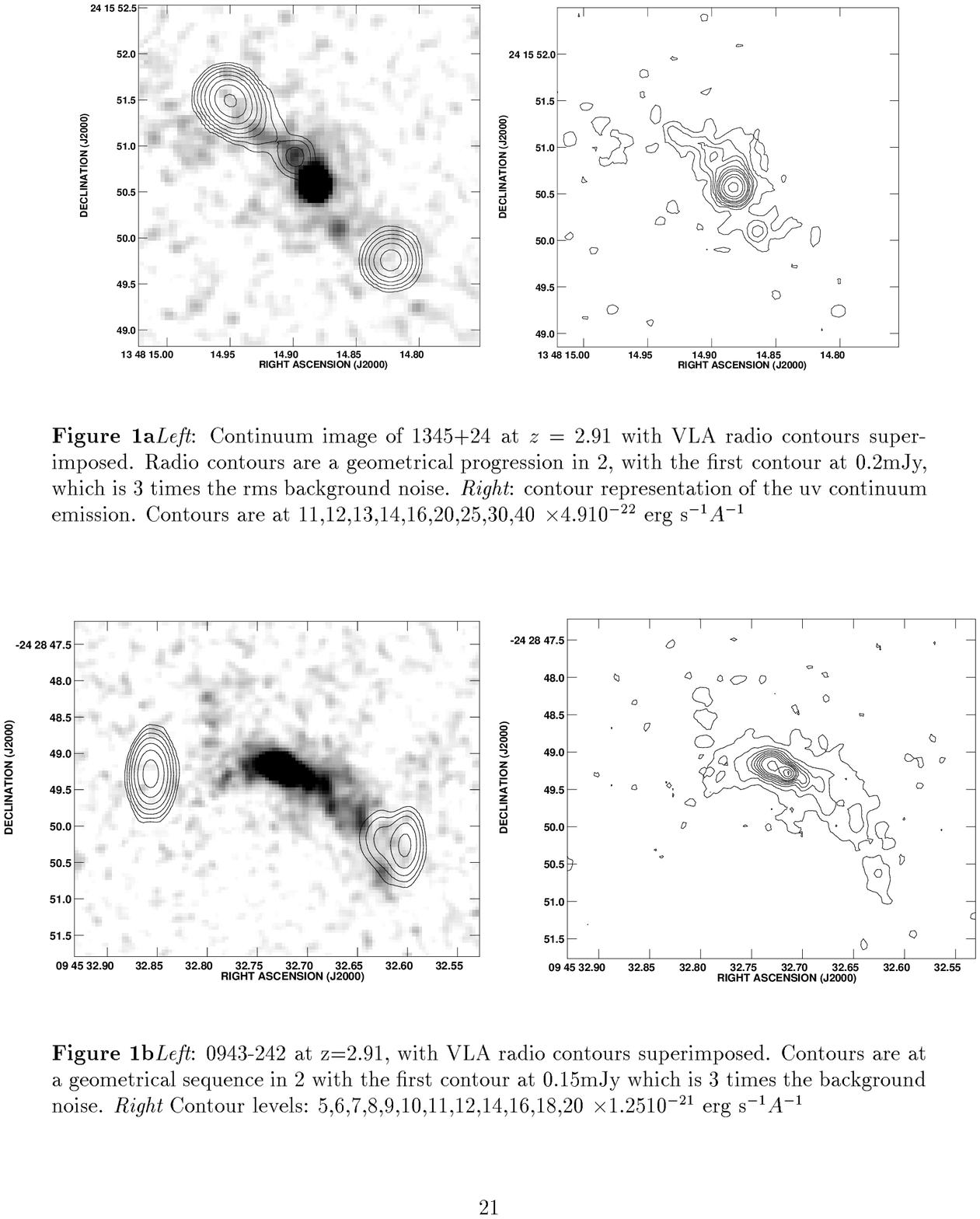,width=8cm,clip=}
\psfig{file=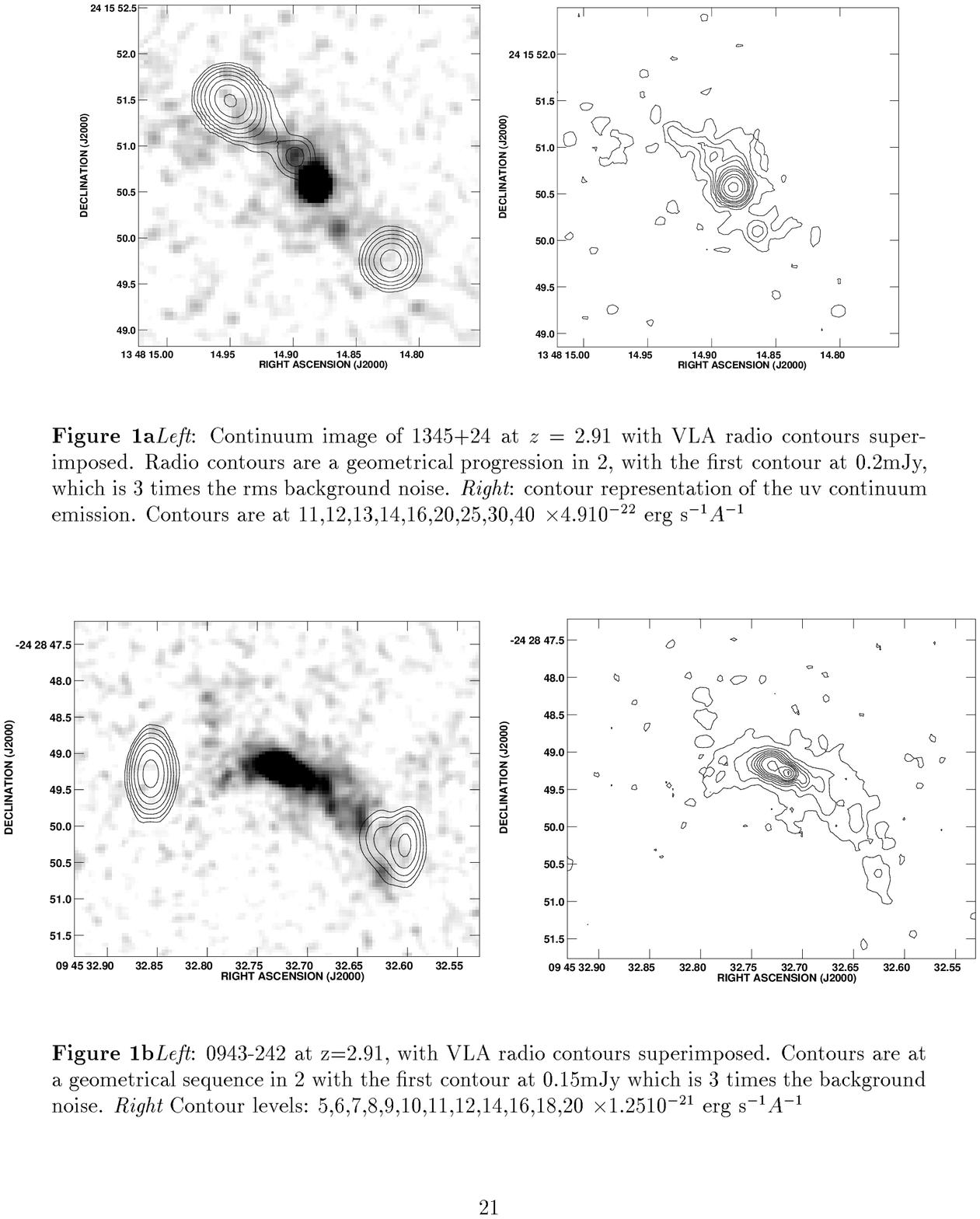,width=8cm,clip=}
}
\caption{
{\it Left}: Grey scale image of MRC 0943$-$242 at z=2.923, with VLA radio contours 
superimposed. Contours are a geometrical  
sequence in steps of 2  with the first contour at 0.15 mJy  
which is 3 times the background noise. {\it Right}: Contour representation of the UV continuum emission. Contour levels are: 5,6,7,8,9,10,11,12,14,16,18,20 $\times 1.25 10^{-21}$ erg s$^{-1}$\AA$^{-1}$.}
\end{figure*}
The new HST image shows that 
in  UV continuum the emission has a  bright compact nucleus. 
On the eastern side of this component there is a jet-like feature 
that follows remarkably well the small curvature of the radio jet: 
this suggests that we might be observing the optical counterpart 
of the radio-jet.
However the radio-to-optical spectral index derived from the flux of the 
component (0.7) is completely different from the high-frequency radio spectral 
index (-1.2). Such flattening of spectral indices into the optical 
is contrary to what is found for sources 
with observed optical synchrotron radiation (e.g. Meisenheimer et al.
1989). \nocite{mei89}
Therefore we discard this possibility. 
A more likely interpretation is that star formation 
is taking place in that region 
triggered by the passage of  the radio jet.
Other possible mechanisms to enhance the emission along the radio jet path
have been proposed by Bremer et al. (1997). \nocite{bre97}  
In Sect. 6.1 we will discuss more extensively the alignment effect
and how all the various models that have been proposed to explain it, 
apply to our sample of radio galaxies.
\\
The object along the radio axis detected by Chambers et al. (see above)
is also detected in our HST image (it is outside the field shown in Fig. 1);
its morphology indicates that it is most probably an edge-on spiral
(hence a foreground object).
\\
\\
{\bf MRC 0943-242}
\\
This radio source at z=2.923 (\cite{rot95a}) 
is only 29 kpc in extent and has a simple double-morphology, with no 
nucleus detected in the present VLA images (\cite{car97}).
The HST image shows a bright elongated main component, 
plus a number of smaller  clumps embedded in a halo of lower surface 
brightness emission with a peculiar  overall
curved morphology.
The inner region of the UV emission shows a remarkably good alignment 
(within  $10^{\circ}$) with the radio axis.
For comparison,the Keck K-band image taken by van Breugel et al. (1998) 
shows a somewhat  rounder and more centrally concentrated morphology. 
\\
High resolution spectroscopy  of the Ly$\alpha$ line shows spatially resolved 
absorption by associated neutral hydrogen, with the absorber 
covering the entire extended Ly$\alpha$ emission (\cite{rot95a}).  
Similar deep high resolution observations have been performed 
 on the CIV and HeII lines: the HeII line does not show absorption, which is expected as this is a resonant line, while the CIV line shows absorption 
due to the CIV 1548/1551 doublet. The column density for the absorber is 
$10^{14.4}$ cm$^{-2}$. Combined with the measured column density 
for  the HI  absorber, this indicates that the spatially extended absorber is metal enriched (\cite{rot96d}).
\\
\\
\begin{figure}
\psfig{file=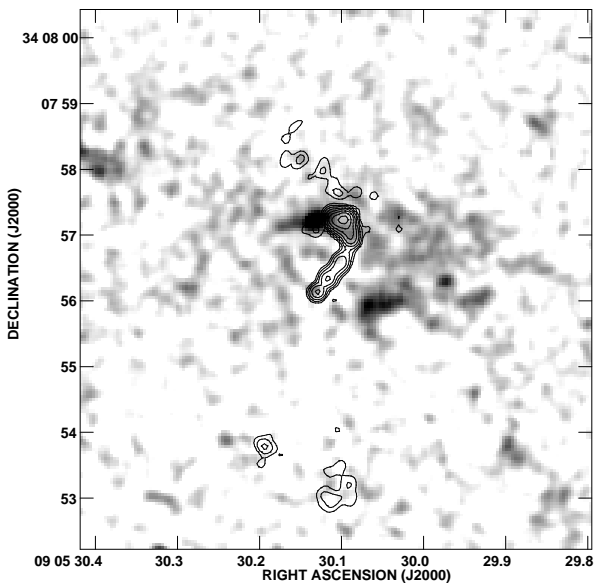,width=8cm,clip=}
\caption{Grey scale representation of the continuum emission from 
B2 0902+343 z=3.395, with VLA contours superimposed.
 Contours are at  geometrical  
sequence in steps of 2 with the first contour at 0.5 mJy.}
\end{figure}
{\bf B2 0902+343}
\\
This radio galaxy was identified by Lilly (1988) \nocite{lil88} and
is one of the most extensively studied high redshift radio 
galaxies. It is 32 kpc in extent.
\\
The radio emission has a bizarre structure showing a bright 
knotty jet with a sharp bend of almost 90$^{\circ}$ at its northern end,
and two southern components whose common orientation is
perpendicular to the rest 
of the source (Carilli et al. 1994).\nocite{car94a}
Further multi-frequency radio studies lead  Carilli (1995) to conclude
that most of the peculiarities of the radio galaxy can be explained  
by assuming that the source is oriented 
at a substantial angle (between 45 and 60 degrees) with respect to 
the plane of the sky, with the  northern regions of the source  approaching 
and that the central region of the galaxy is obscured by a 
substantial amount of dust.
\\
From extensive studies Eisenhardt \& Dickinson (1992) \nocite{eis92} 
found that B2 0902+343  has  a flat optical spectral energy distribution (SED),
and an unusually low surface brightness
  distribution at optical and IR wavelengths; this 
lead to the suggestion that B2 0902+343 might be a 
proto-galaxy, undergoing a first major burst of star formation 
(Eales et al. 1993, Eisenhardt \& Dickinson 1992).\nocite{eal93b}
The presence of associated 21 cm neutral hydrogen in absorption 
against the radio continuum source was 
first detected by by Uson et al. (1991) and confirmed 
by others (\cite{bri93,bru96}). \nocite{uso91} 
However no strong absorption in the Ly$\alpha$ emission line has 
 been detected (\cite{mar95}).
\\
The optical morphology, as imaged by the HST, 
confirms the unusually low surface brightness distribution and shows
that the galaxy  consists of 2 regions,
of approximately the same flux with a void in between, plus an extended 
fuzzy  emission region to the north east of them. The source does not
exhibit the radio-optical alignment effect; the UV emission is almost 
perpendicular to the radio axis. 
With the present astrometry the radio core is situated in a 
valley between the optical peaks; this morphology could be explained
with the presence of large amounts of dusts.
However the uncertainties 
in the astrometry are such that the radio core could be 
coincident with any of the two optical components. 
\\
\\
{\bf MRC 2025-218}
\\
The galaxy associated with this USS radio source at z=2.630 (38 kpc in 
extent), was first 
identified by McCarthy et al. (1990). \nocite{mcc90a}
Deep multi-frequency radio imaging show a double radio source 
with a jet on the southern side of the core, which  
has an extremely sharp bend towards the west, making an angle 
of $\sim 90^{\circ}$  (\cite{car97}). The northern lobe has a faint
extension in the direction of the core which could be a counter--jet.  
Ground--based 
near infrared imaging show a compact object (van Breugel et al. 1998),
while the Ly$\alpha$ 
emission extends for more than 5$''$ along the radio axis and is distributed bimodally. 
The total SED of the galaxy is well fit by a main 
stellar population aged 1.5 Gyrs, combined with 
a young star-burst contributing
20\% of the total light at 5000 \AA \space(\cite{mcc92a}).
Cimatti et al. (1993) find that that the rest frame UV continuum 
emission is linearly
 polarized ($P=8.3 \pm 2.3 \%$), with the electric vector oriented 
perpendicular to the UV emission axis. 
\\
The HST image shows that the  host galaxy has 
a compact morphology, consisting of a bright nucleus, 
two smaller components and extended low surface brightness emission,
which is  elongated and well aligned with the radio axis. 
The angle between the inner radio axis and the extended UV emission 
is only $\sim 5^{\circ} \pm 3^{\circ}$. 
There is  no direct one-to-one relation 
between the radio components and the UV emission, unlike 4C 1345+245;
however if we draw a cone of opening angle $\sim 30^{\circ}$ along
the radio axis,   
all the UV emission on both sides of the radio core is  then constrained
within this cone. 
Such a morphology, reminiscent of an ionization cone, is expected 
in  models where the aligned UV continuum
emission is scattered light of a buried quasar, and is
supported by the polarization measurements by  Cimatti et al. (1993).
\\
The present HST image reveals little UV emission 
near the bend: however 
high resolution  spectroscopic observations of the Ly$\alpha$ 
emission line  show that the galaxy is embedded in a very large halo of
ionized gas, extended well beyond  the radio source 
(more than 60 kpc i.e. double the size of the radio source); therefore 
the most likely explanation for the bend is that  interaction 
between the radio plasma and the surrounding gas deflects the jet, 
as observed in other cases
(e.g. Pentericci et al. 1997).
\\
\\
\begin{figure*}
\centerline{
\psfig{file=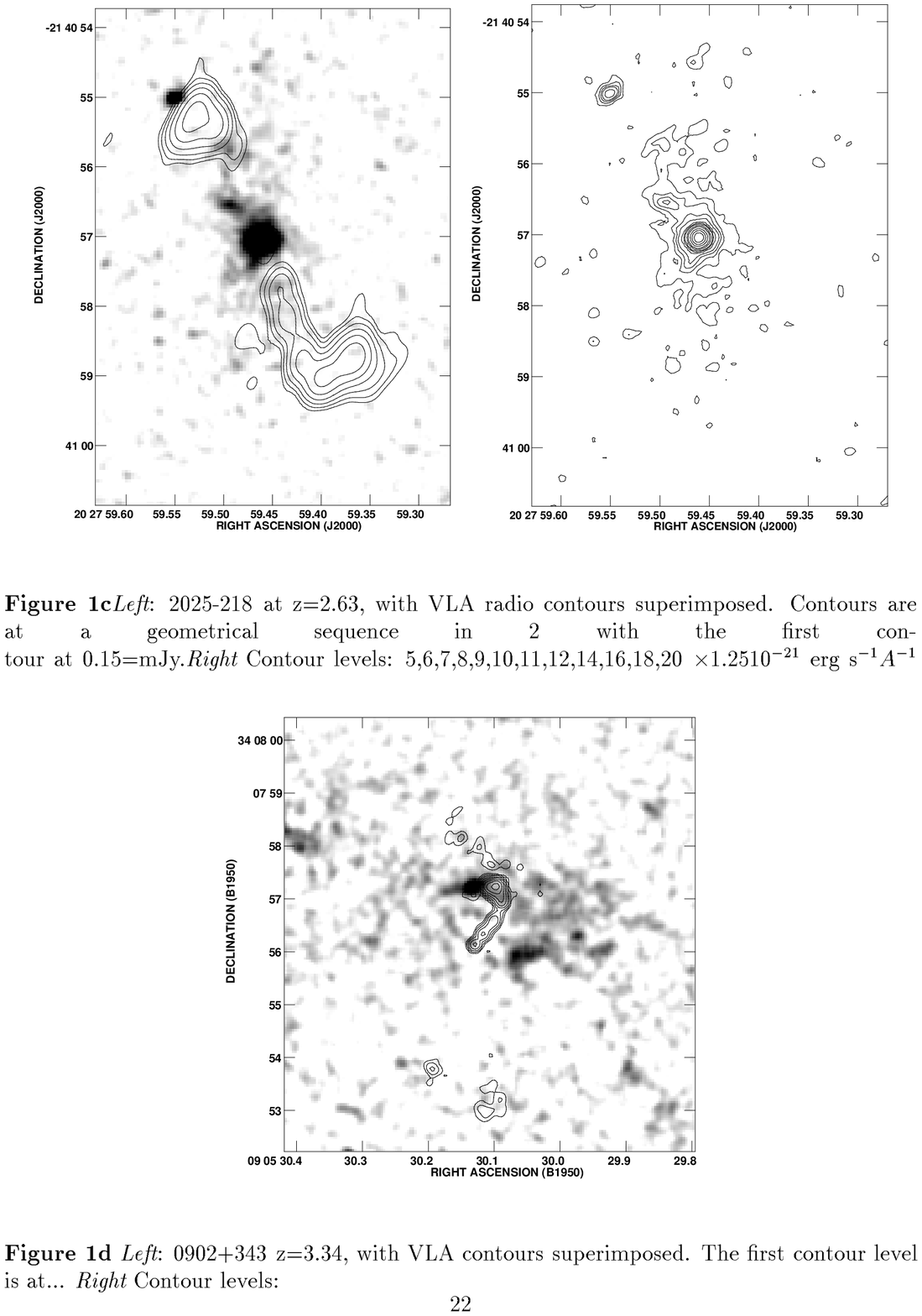,width=8cm,clip=}
\psfig{file=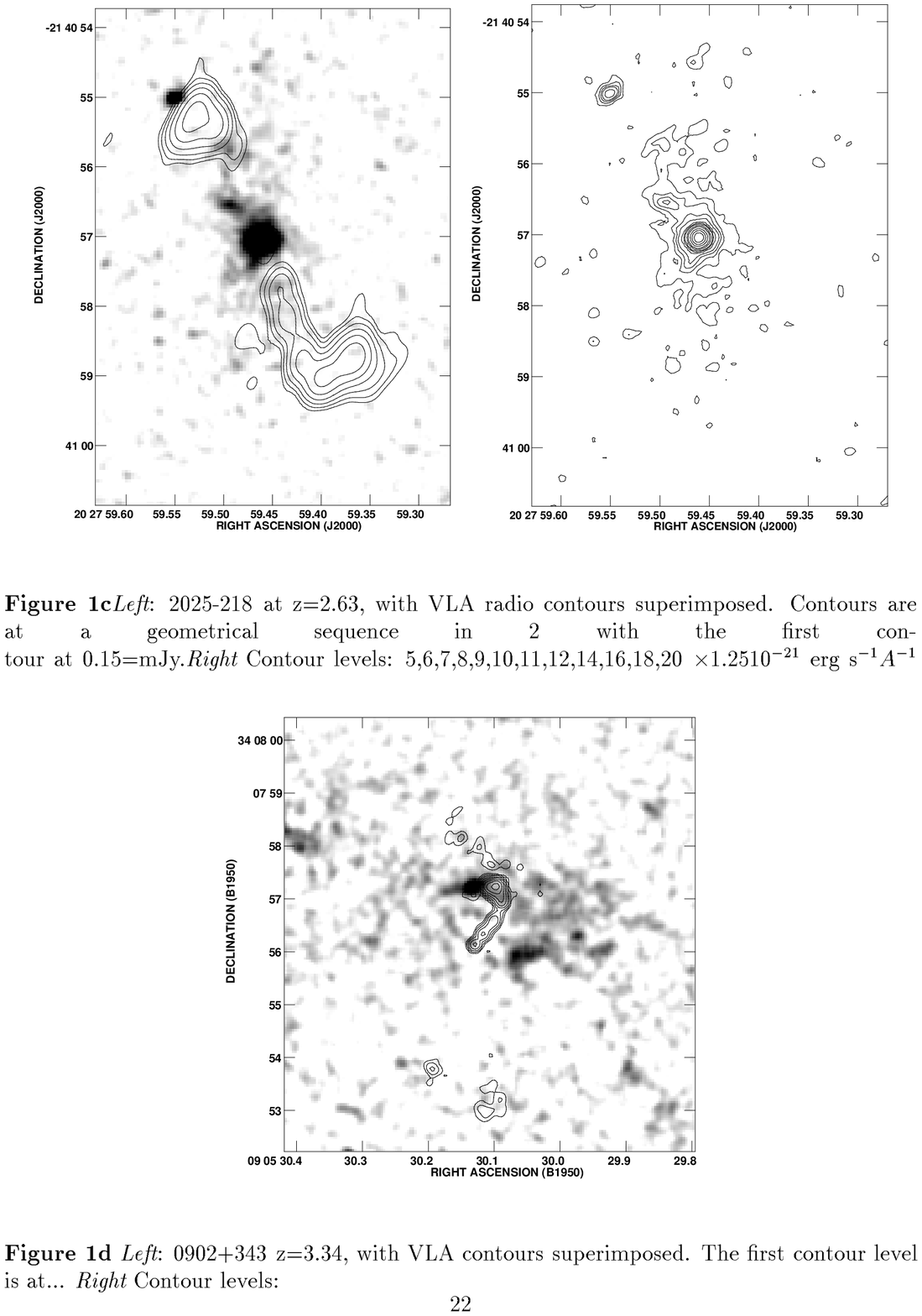,width=8cm,clip=}
}
\caption{{\it Left}: Grey scale image of MRC 2025$-$218 at z=2.630, 
with VLA radio contours superimposed. Contours are a geometrical sequence 
in steps of 2 with the first contour at 0.15 mJy. {\it Right}: Contour representation of the UV continuum emission. Contour levels: 5,6,7,8,9,10,11,12,14,16,18,20 $\times 1.25 10^{-21}$ erg s$^{-1}$\AA$^{-1}$.}
\end{figure*}
{\bf  4C 1243+036} 
\\
This radio galaxy at z=3.570 which has an extension of 50 kpc, 
was identified and 
extensively studied by van Ojik et al. (1996).
The radio source is  double  with a sharp bent structure on the 
southern side. Strong depolarization of the radio emission indicates 
that the source is embedded in a magneto-ionic medium.
High resolution spectroscopy and narrow band imaging of the Ly$\alpha$ 
emission line have detected the presence of a giant (100 kpc) halo
of ionized gas showing ordered motion, possibly due to rotation 
of a proto-galactic gas disk, out of which the galaxy 
associated with 4C 1243+036 is forming.
Furthermore the Ly$\alpha$ emission shows a secondary peak at the location
of the bending of the radio jet, consistent with a
 gas cloud being  responsible for the deflection of the radio 
jet (Ojik et al. 1996).
\\
The morphology of the galaxy as imaged by the HST consists 
of a nucleus from where a narrow and  elongated structure departs, which
then bends to the south. There is also a 
smaller component, about 1$''$ beyond the northern radio hot-spot, 
which could  belong to the system, since narrow--band Ly$\alpha$ 
imaging shows that
there is Ly$\alpha$ emission at this location (van Ojik et al. 1996).\nocite{oji96}
\\
The most remarkable  characteristic of 4C 1243+036 is that the UV light 
follows closely the direction of the radio source, both in the 
inner 2$''$ region where the light is aligned with the radio axis 
to within 15 degrees, but especially at the location of the bend: 
here both the UV emission and the radio jet bend rather sharply 
to the south, suggesting a direct relation between the radio jet 
and the UV component. This is similar to 
the case of  the radio galaxy 4C 1345+245.
Note that recent K-band Keck imaging of 4C 1243+036 by van Breugel 
 et al. (1998), although at a different resolution, 
indicate that also the K-band continuum emission is elongated and follows 
the bend of the radio jet.
\\
\\
{\bf MG 2141+192}
\\
This galaxy at z=3.594 (60 kpc in extent) was identified by Spinrad  
et al. (1992) and since then has been extensively studied by various 
groups.
The radio source has a simple double morphology, with no nucleus  
detected in the present  images. 
Eales \& Rawlings (1996)\nocite{eal96}
who imaged this object in the infrared, report the detection of a 
 relatively brighter component
half way between the radio hot-spots and a second fainter one, 
$4''$ north,  approximately coincident with the northern radio hot-spot.
Recently van Breugel et al. (1998) re-imaged the object in 
the near-infrared with the Keck telescope, 
finding  additional extended low surface brightness 
emission southern of the nucleus.
Armus et al. (1998) imaged 
the [OIII] emission line nebula associated with the galaxy, which has an extent
of more than 70 kpc (equal to the separation between  the radio lobes), 
is extremely narrow and aligned with the radio axis. 
By comparing fluxes of the different emission lines 
they also find indications for the presence of large amounts of 
dust.\nocite{arm98} Finally Maxfield et al. (1997) find that the 
emission nebulae of Ly$\alpha$, CIV and HeII are not only spatially 
extended  but also have remarkable velocity structure 
with multi-components velocity displacement up to 1900 Km s$^{-1}$,  
which are most consistent with a shock ionization picture. 
\\
The HST shows that the host galaxy is very faint in the UV restframe, 
and consists of a nucleus with a faint filamentary extension 
and a small clump to the west.
In the HST image  some fuzzy emission (at a 3$\sigma$ 
level) is present near the position of the radio hot-spot, 
where the second infrared
component is located. 
We also detect similar
 emission very close to the position of the southern radio component.
Overall the UV rest-frame emission is extremely faint, consistent
with the presence of large amounts of dust.
Deeper images are needed to delineate the morphology of this galaxy
in more detail.
\\
\\
{\bf TX 0828+193 }
\\
This large radio source (98 kpc in extent) at z=2.572 (van Ojik 1995), 
has a double morphology
with a jet extending  from the core towards the northern hot-spot 
(most probably the approaching side). The end of the jet contains  
multiple hot-spots and has a 90 degrees bent. 
The southern part of the radio source consists only of a single hot-spot.
\\  
The HST image shows a small galaxy consisting  of several clumps arranged in a
triangular shape.
We choose to identify the radio core with the brightest optical component
(the same procedure we followed for other galaxies, see Sect. 4); 
however another possible registration would be 
with the radio core at the vertex of the triangulum.
At this position, an ionization cone on both sides
of the radio core, would encompass all the UV emission.
The morphology of TX 0828+193, 
like that of MRC 2025$-$218, strongly suggest that a large fraction of the
UV light might be 
scattered light from a buried AGN. 
The axis of this ``scattering cone'' is aligned with the radio 
axis to within a few degrees 
($7^{\circ} \pm 3^{\circ}$).
\\
There is another object located 
along the radio axis which could be associated with
the radio source (a companion galaxy): it is bright in the UV continuum but shows no line emission, so it could as well be an intervening 
system at a different redshift (\cite{oji97a}).
\\
The Ly$\alpha$ emission from this radio galaxy 
has a spectacular shape, with the 
entire blue wing of the emission line profile absorbed 
by neutral gas associated with the galaxy 
(van Ojik et al. 1997). If the companion object is at the same redshift 
as TX 0828+193, then it is  possible that a neutral gaseous halo associated 
with it is the responsible for the absorption.
Since the absorption is very steep and broad, it is probably due
to a combination of absorbing systems each at a slightly different velocity 
with respect to the Ly$\alpha$ peak. Also in the red wing of the Ly$\alpha$ profile a broad shoulder is observed that maybe be due to multiple HI absorption
systems or to intrinsic velocity structure in the ionized gas.
\\
\\
{\bf TX 0211-122 }
\\
This large radio source (134 kpc) at z=2.336 (van Ojik et al., 1994) 
has a simple double morphology.  
A jet feature extends from the core towards south, 
curves and reaches the eastern lobe; this 
structure suggests that the radio axis might be precessing.
\\
The galaxy, as shown from the HST image, 
consists of a  bright nucleus and a much smaller clump, both embedded 
by  lower surface brightness emission, distributed in an irregular way.
The contour image of the central component shows that it consists of 
two ``tails'', one of which points in the direction of the inner radio jet.
\\
The optical spectrum of this source is peculiar with  the  
Ly$\alpha$ emission being anomalously weak when compared to 
higher ionization lines: the flux ratio of 
Ly$\alpha$ to NV a factor of 30 smaller than that of typical HZRGs 
while the large NV/CIV ratio indicates that the line-emitting gas 
is over-abundant in nitrogen (\cite{oji94a}). 
Van Ojik et al. consider various mechanism that could produce these features,
and conclude that the galaxy is likely to  be undergoing a massive star-burst in the central region, possibly as the result of the passage of the radio jet. 
The star-burst would produce large amounts of dust, which when mixed 
through the emission line gas partly absorbs the Ly$\alpha$ emission, giving 
it a very patchy morphology, while the enhancement of nitrogen emission could be produced either by shocks or photo-ionization. 
\\
\\
{\bf TX 1707+105}
\\
This radio source at z=2.349 (van Ojik 1995) which is 173 kpc in extent 
is one of the most peculiar systems in  our sample: 
it consists of two galaxies (labeled A and B in Fig. 9) both 
showing  strong and extended Ly$\alpha$ emission at the same redshift. 
The 2 objects lie almost exactly along the radio axis 
and they are both clumpy 
and elongated in a direction with is almost perpendicular to it.
In particular galaxy 1707A (the brightest one)  
is comprised of a series of knots
of approximately the same brightness, which form a sort of string, while galaxy 1707B consists of only two clumps.
There is a further emission component, which in Fig. 9 is indicated 
as C, that 
lies in between the two galaxies and could be part of the system. 
It does not show line emission, although probably when the high resolution 
spectrum was taken
this object fell outside our $2''$ wide slit.
\\
With the present data it is not possible to determine exactly 
which galaxy is associated with the radio emission. 
Given the large extension of the source,  
we expect, for symmetry reasons, that the radio source is associated to the 
galaxy closest to the  center, i.e. galaxy 1707A.
If this is the case then 1707B, and possibly 1707C, would be companion 
galaxies located along the radio axis. There are many cases of companion
galaxies of high and low redshift radio galaxies.
The best known case is Minkowsky's object: 
the location of this  dwarf galaxy is at the end of the radio jet emanating from the radio galaxy PKS 0123$-$016 at $ z=0.0181$, suggesting that its  origin 
is due to jet-induced star formation (\cite{bre85c}). A similar star forming region associated with the nearby powerful radio galaxy 3C285 has also been reported (\cite{bre93b}).
The most recent example is the radio source 
3C34 (at z=0.69), which shows a clumpy emission feature
 along the radio axis and and oriented towards a radio hot-spot. 
Also in this case, the emission has been associated with a region of massive 
star formation triggered by the passage of the radio jet (\cite{bes97}).
Finally, R{\"o}ttgering et al. (1996) find that companion galaxies 
of radio sources tend to be distributed along the direction of the radio axis, which, in their interpretation, could be due 
 to the luminosity of merging dwarf galaxies being enhanced by 
scattering and/or jet-induced star formation.\nocite{rot96e}
\\
\\
{\bf MRC 2104-242 }  
\\
This radio galaxy at z=2.491 is 177 kpc in extent and was first 
identified by McCarthy et al. 
(1990). \nocite{mcc90a}
It has a simple double morphology 
and a relatively bright nucleus.
\\
\begin{figure*}
\centerline{
\psfig{file=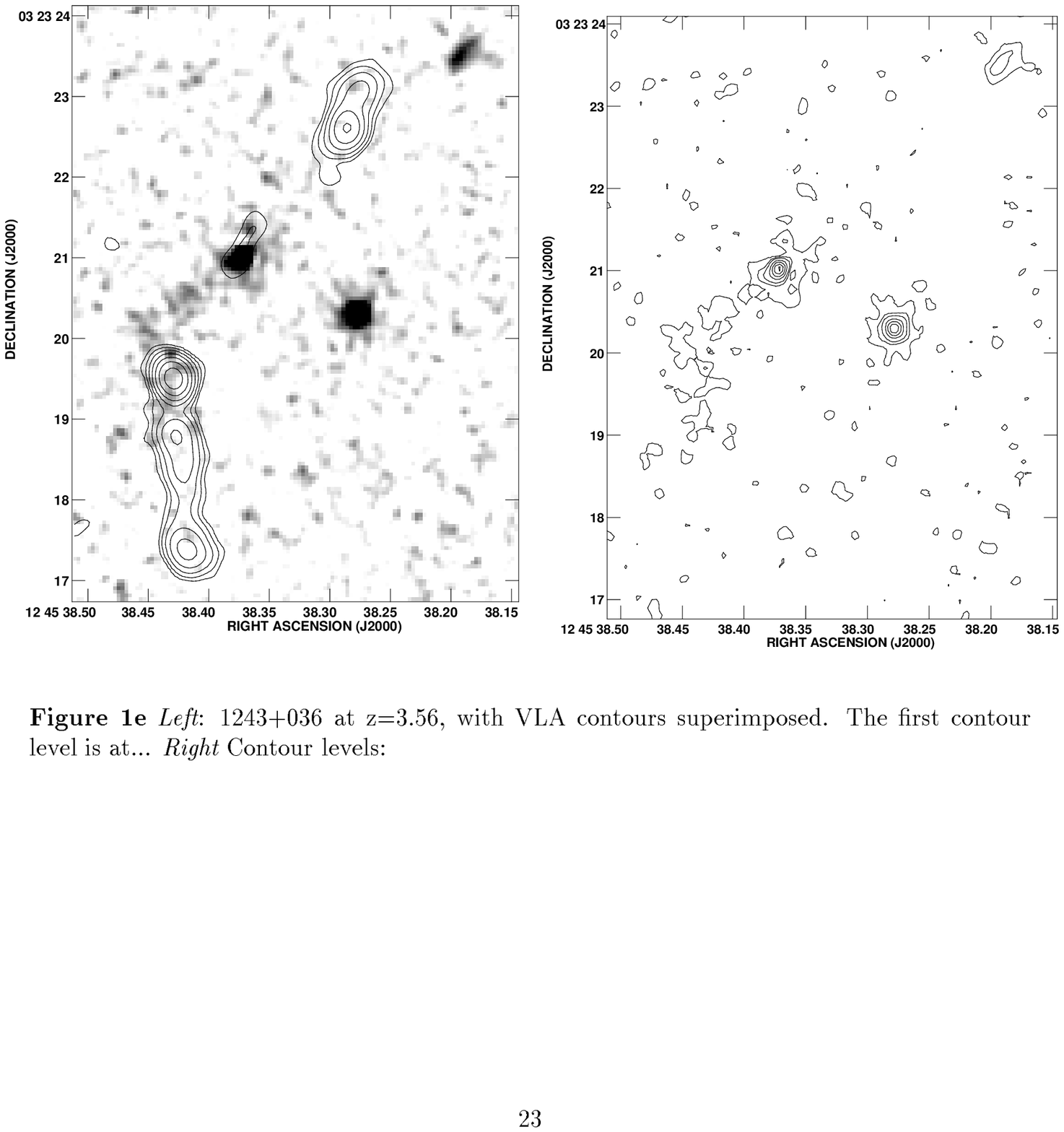,width=7cm,clip=}
\psfig{file=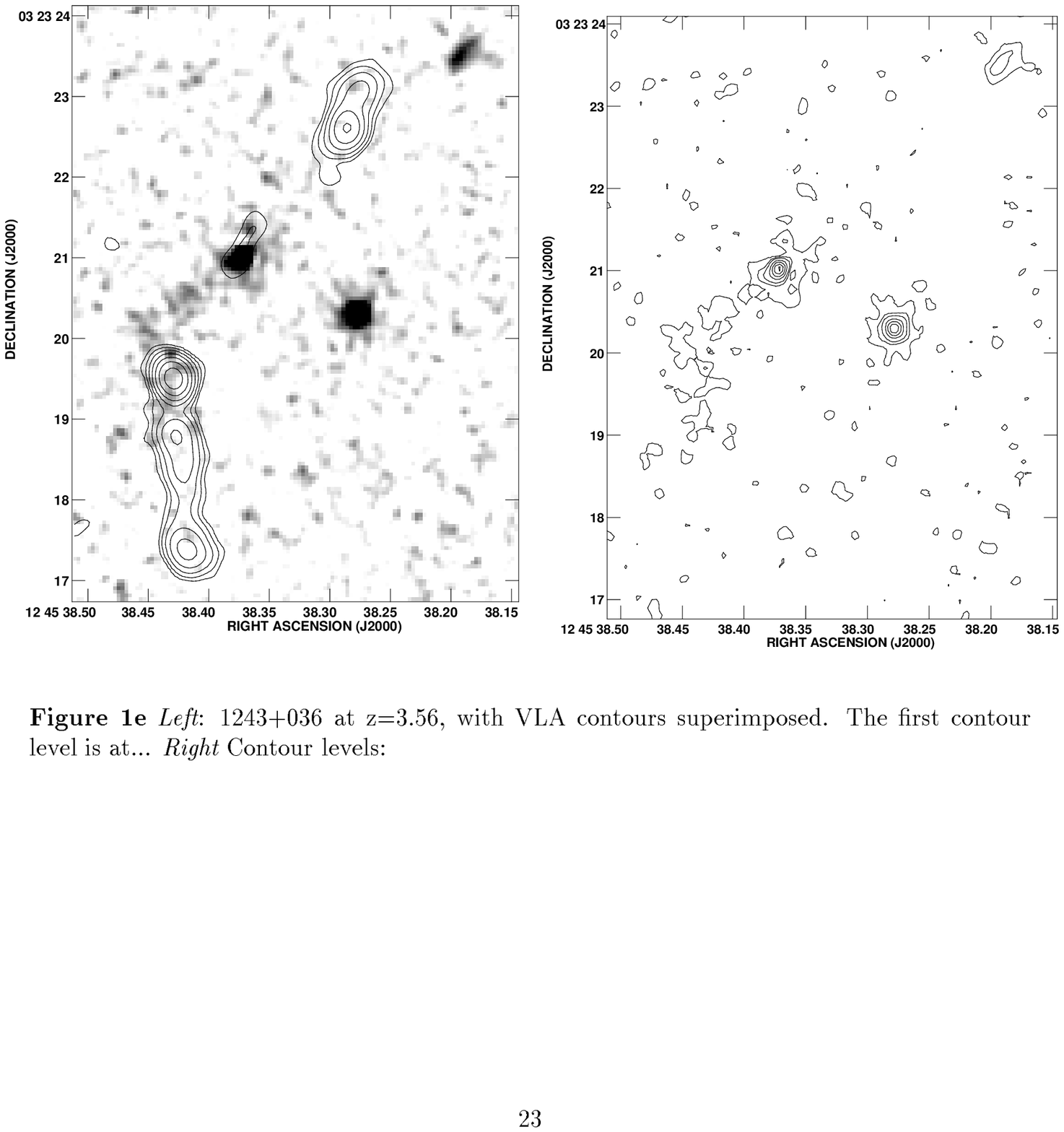,width=7cm,clip=}
}
\caption{{\it Left}: Grey scale image of 4C 1243+036 at z=3.570, 
with VLA contours superimposed.  Contours are a geometrical sequence 
in steps of 2 with the first contour at 0.1 mJy.
{\it Right}:  Contour representation of the UV continuum emission. Contour levels:20,22,24,26,28,30 $\times 9 10^{-22}$ erg s$^{-1}$\AA$^{-1}$.}
\end{figure*}
\begin{figure}
\psfig{file=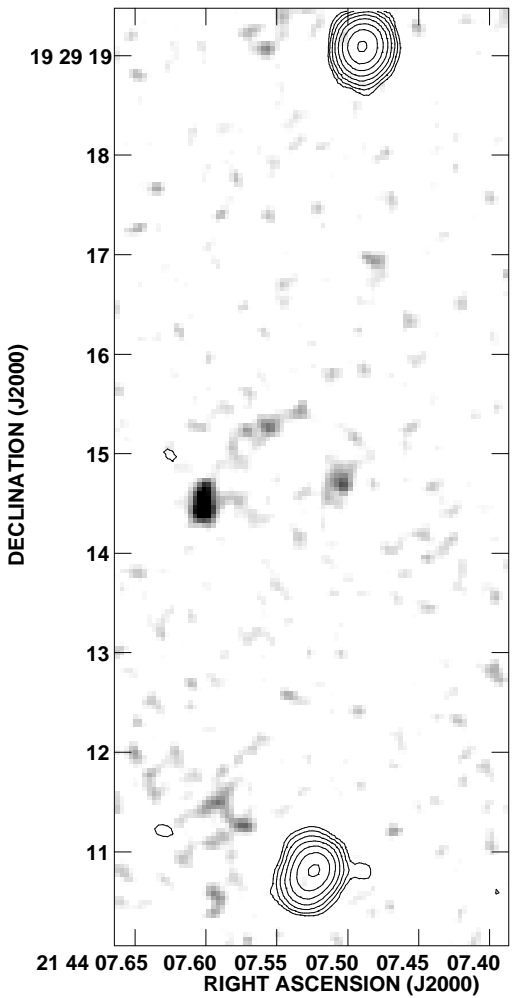,width=7cm,clip=}
\caption{Grey scale representation of the continuum emission of 
MG 2141+192 at z=3.594, with  VLA radio contours 
superimposed. Contours are a geometrical sequence in steps of 
2 with the first contour at 0.1 mJy.}
\end{figure}
\begin{figure*}
\centerline{
\psfig{file=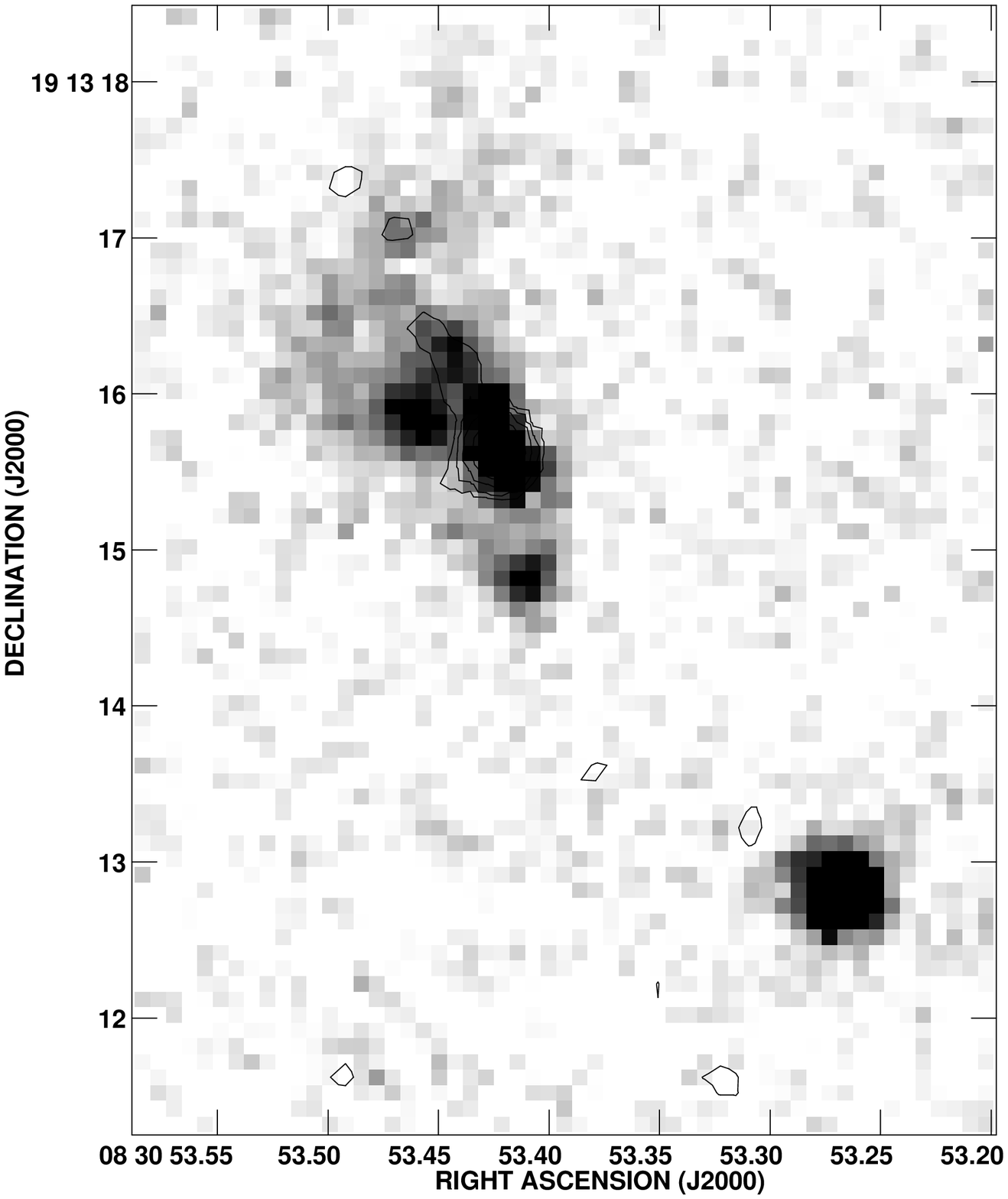,width=7cm,clip=}
\psfig{file=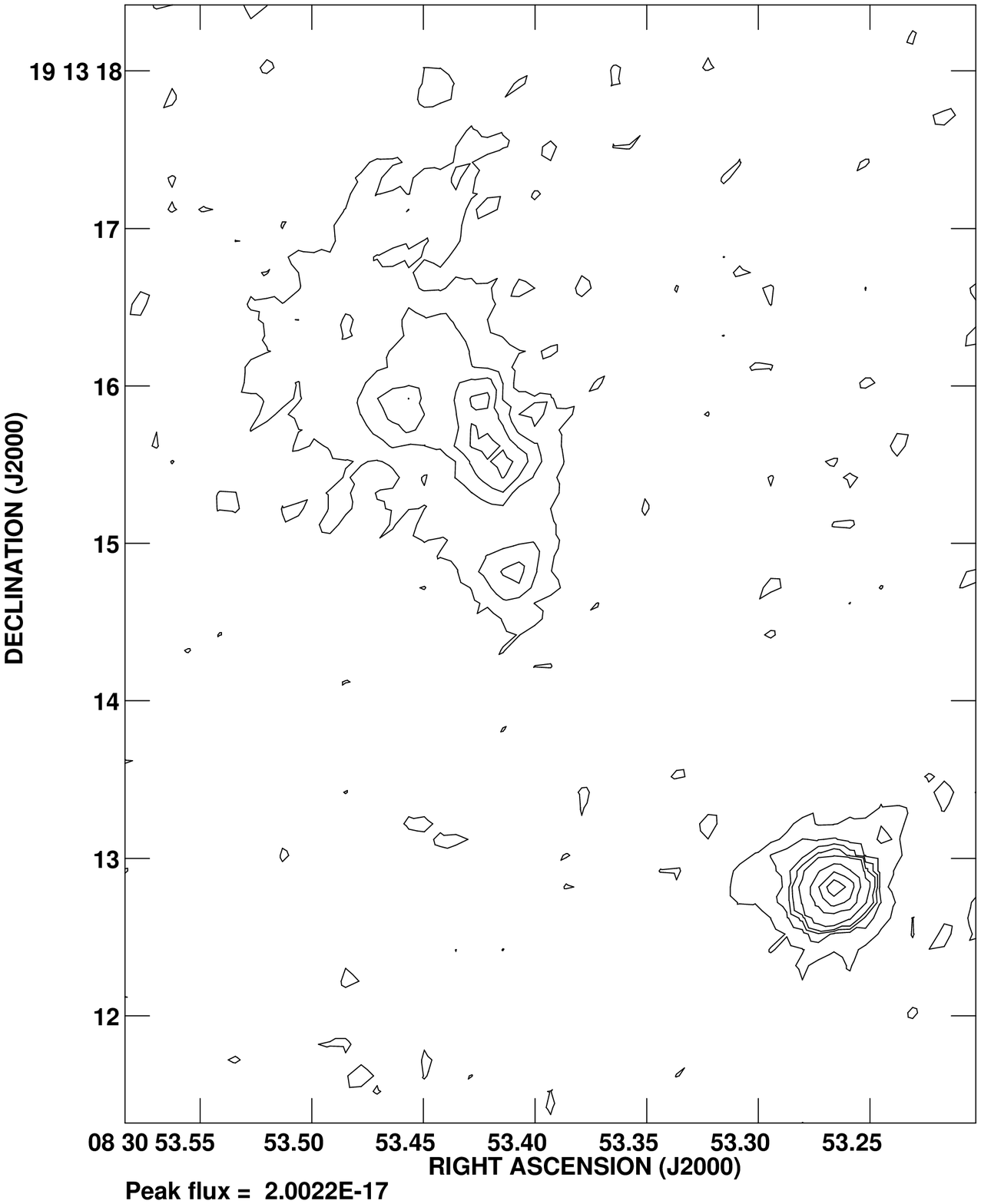,width=7cm,clip=}
}
\caption{{\it Left} Grey scale image of TX 0828+193 at z=2.572, 
with VLA contours superimposed. 
Contours are a geometrical sequence in steps of 
2 with the first contour at 0.11mJy.
{\it Right}:
 Contour representation of the UV continuum emission. Contour levels: 10,12,14,16,18,20,40,80 $\times 4.5 10^{-21}$ erg s$^{-1}$\AA$^{-1}$.
}
\end{figure*}
The Ly$\alpha$ emission is spectacular: narrow band imaging show 
two large gas clumps,  extending  for more
than 12$''$  along the radio axis. Spectroscopy of the line showed
that both components have very large 
velocity distribution ($\sim $1000-1500 km s$^{-1}$), 
large equivalent width and 
have a net velocity difference of about 500 km s$^{-1}$.
Each component contains multiple velocity peaks and kinematic data 
at various position angles indicate that there is no overall ordered motion
(\cite{mcc90a,koe96}).    
A detailed study of the Ly$\alpha$ emission line 
showed that a model based on shocks 
from direct interaction between the radio plasma and the gas can explain 
both the kinematics and the morphology of the gas (\cite{koe96}).
\\
The HST image is remarkable: the 
host galaxy is one of the clumpiest  of our sample,
consisting of a number of knots of similar brightness and size,
located around the radio core. Unfortunately some of the components
are confused with the residuals from a spike
of an extremely bright nearby star. 
Furthermore there is a filamentary  
component that is more than 2$''$ long and extremely narrow.
This last component is aligned with 
the radio axis to within a few degrees.  
The overall extension of the host galaxy is almost 7$''$, 
making it the largest optical galaxy in our sample.
\\
\\
\begin{figure*}
\centerline{
\psfig{file=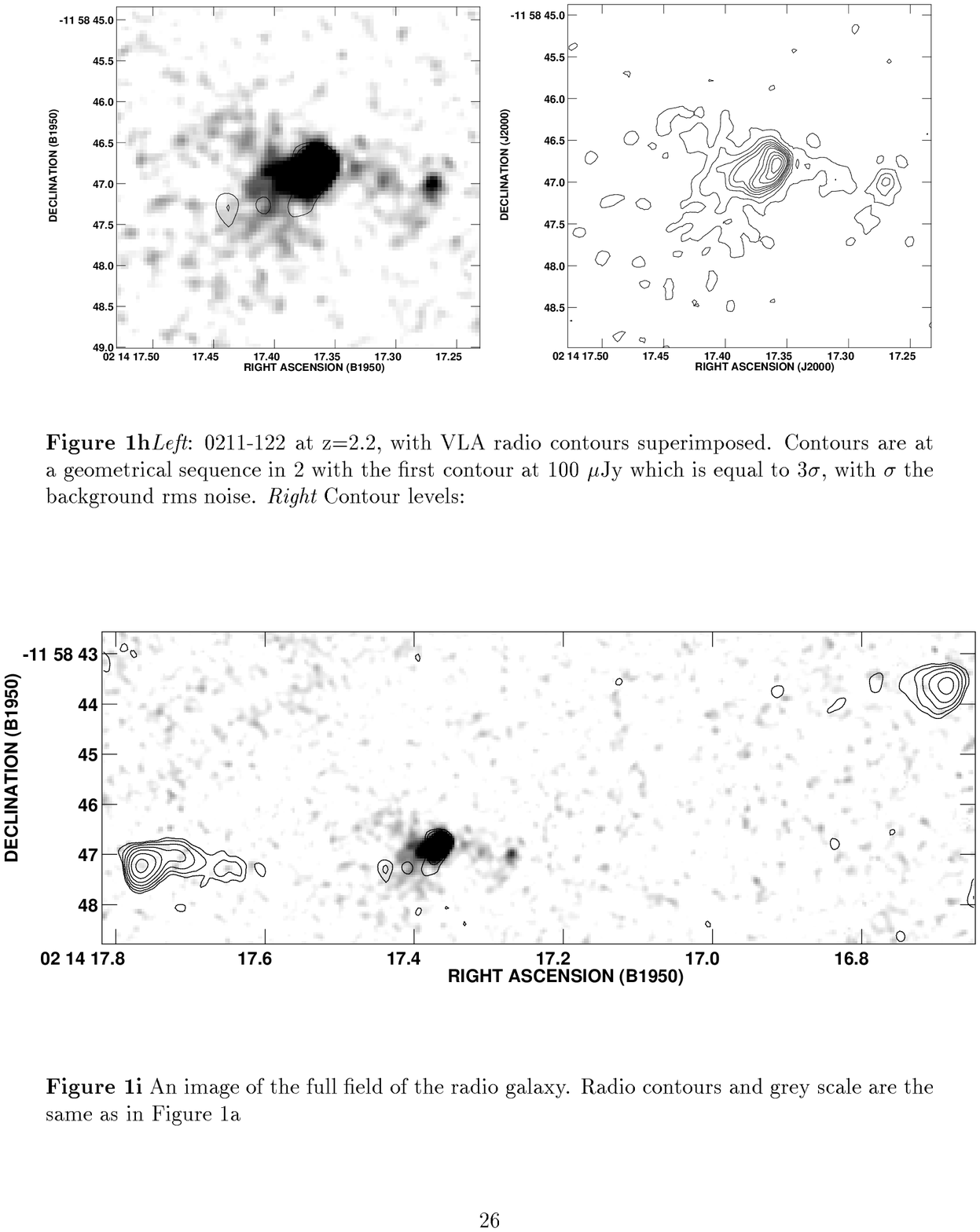,width=7cm,clip=}
\psfig{file=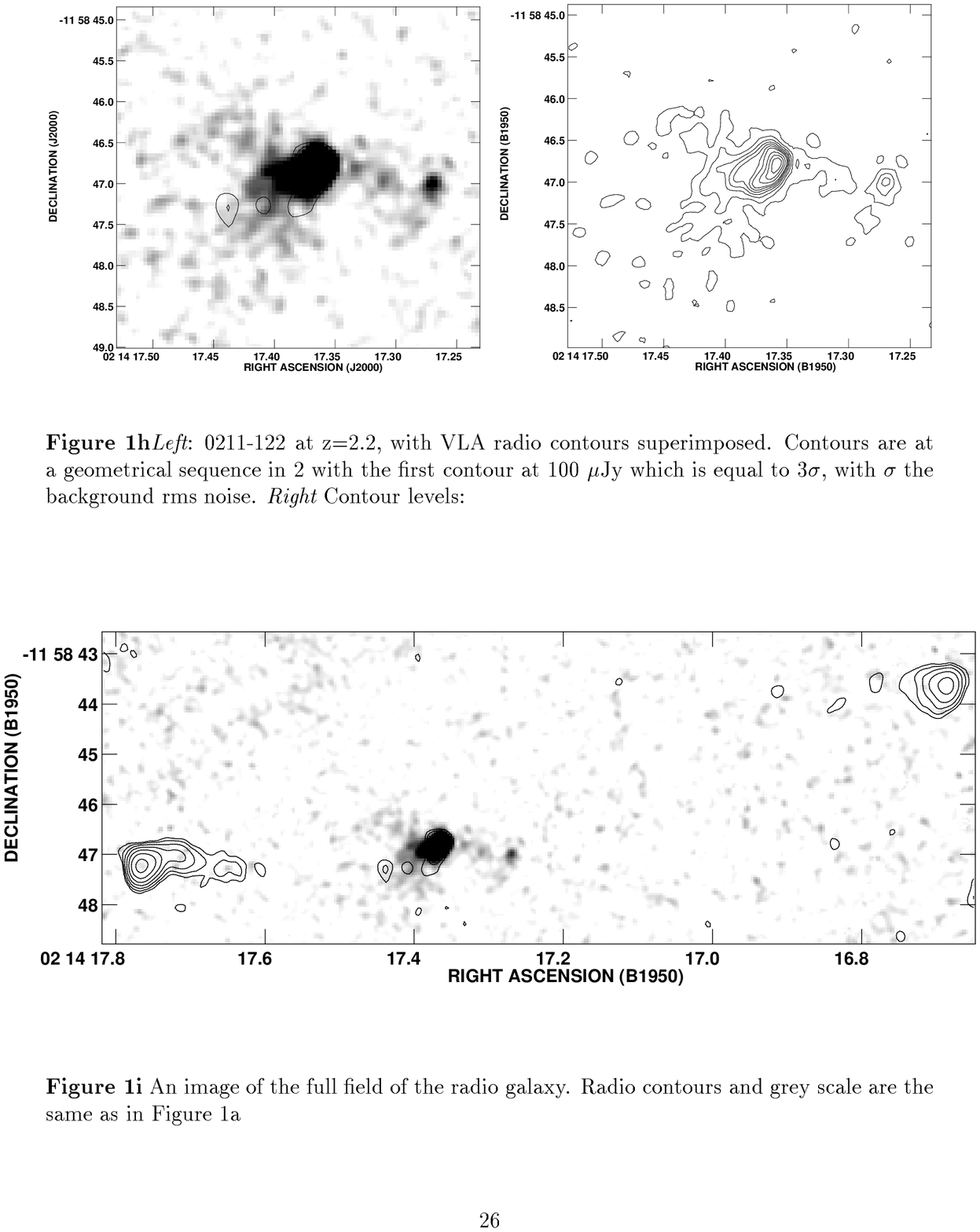,width=7cm,clip=}
}
\centerline{
\psfig{file=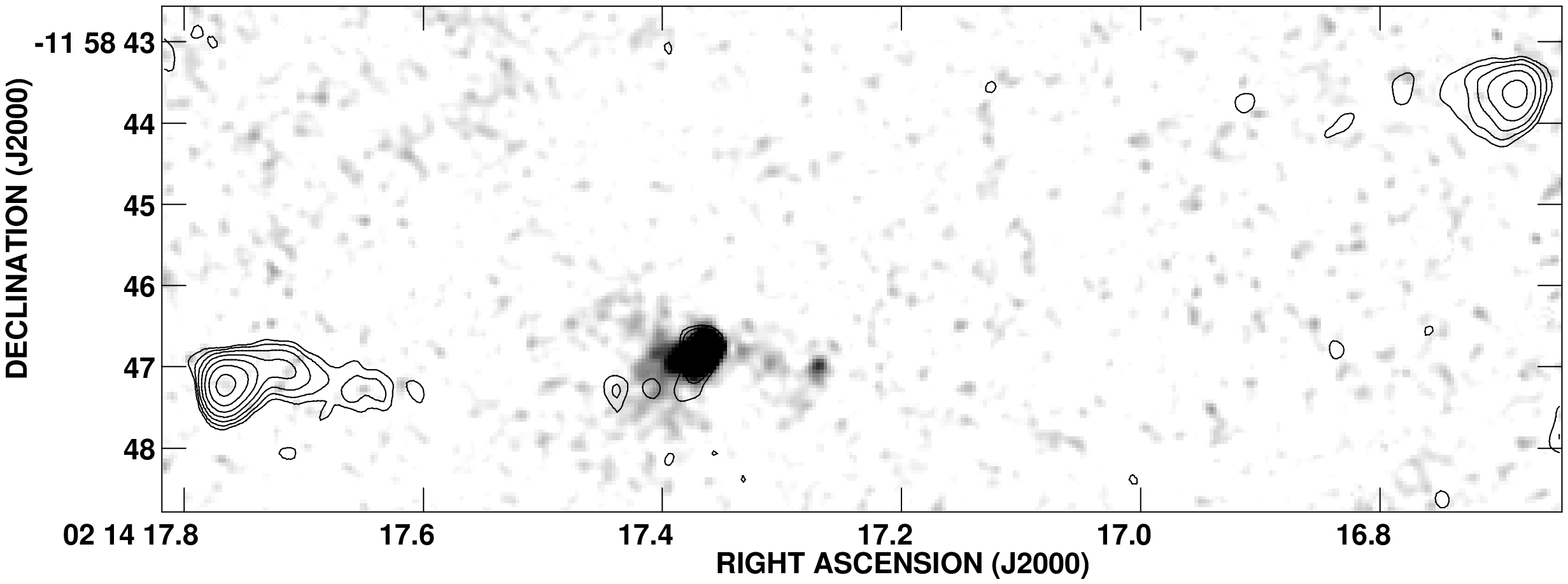,width=18.cm,angle=-90}
} 
\vskip-1cm
\caption{{\it Top left}: Grey scale image of
TX 0211-122 at z=2.336, with VLA radio contours superimposed.
Contours are at a geometrical sequence in 2 with the first contour at 
0.1 mJy. {\it Top right}:  Contour representation of the UV continuum emission. Contour levels:8,9,10,12,14,16,18,20 $\times  10^{-21}$ erg s$^{-1}$\AA$^{-1}$. 
{\it Bottom}: An image of the full field of the radio galaxy. 
Radio contours and grey scale are the same as the upper left panel.}
\end{figure*}
{\bf 4C 1410-001}
\\
This radio galaxy at z=2.363 (van Ojik, 1995) is, with 189 kpc, 
the largest radio source in the sample. 
The host galaxy  
is highly elongated ($\simeq 5''$). 
\begin{figure*}
\centerline{
\psfig{file=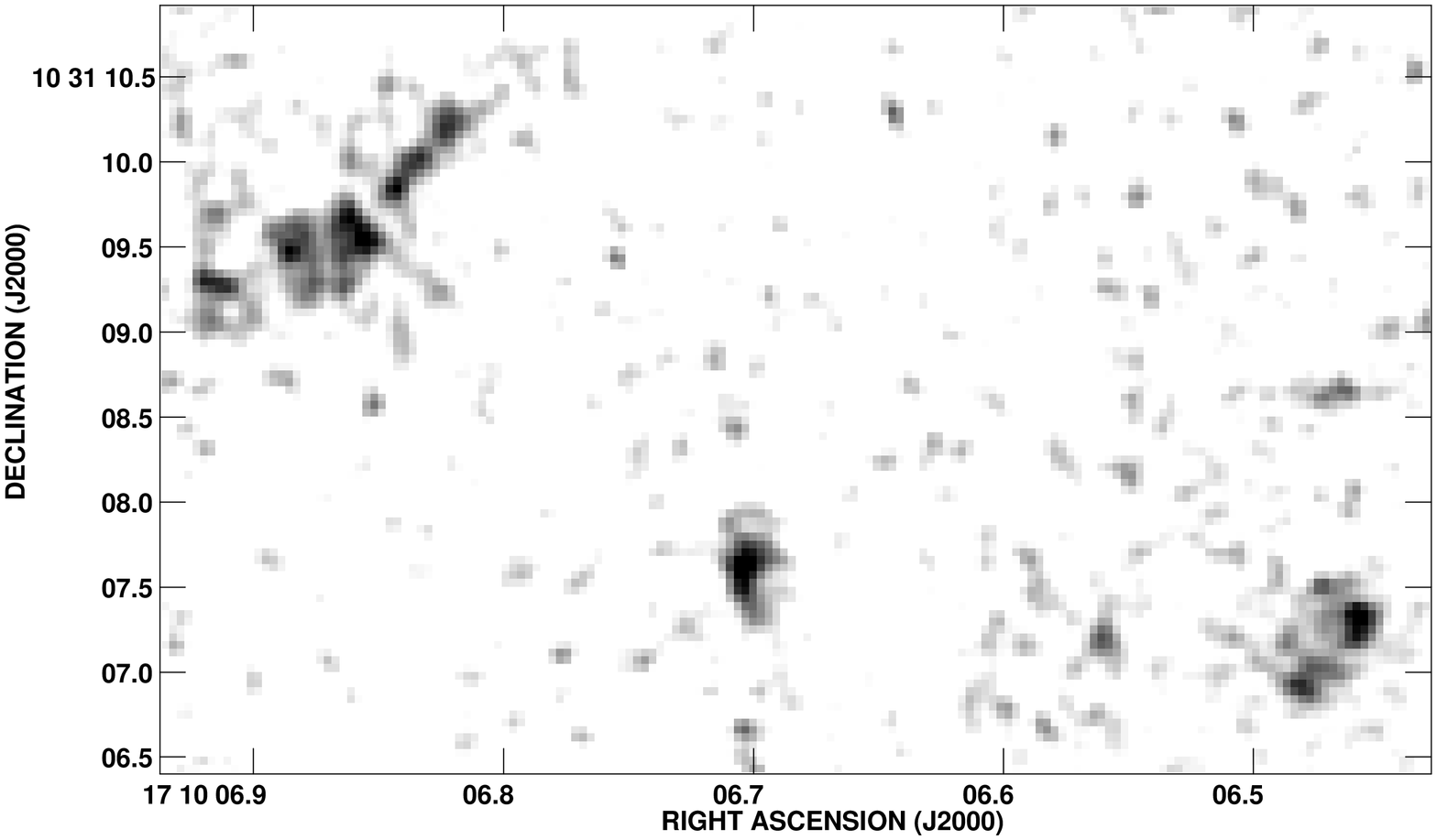,width=8.5cm,angle=-90}
}
\centerline{
\psfig{file=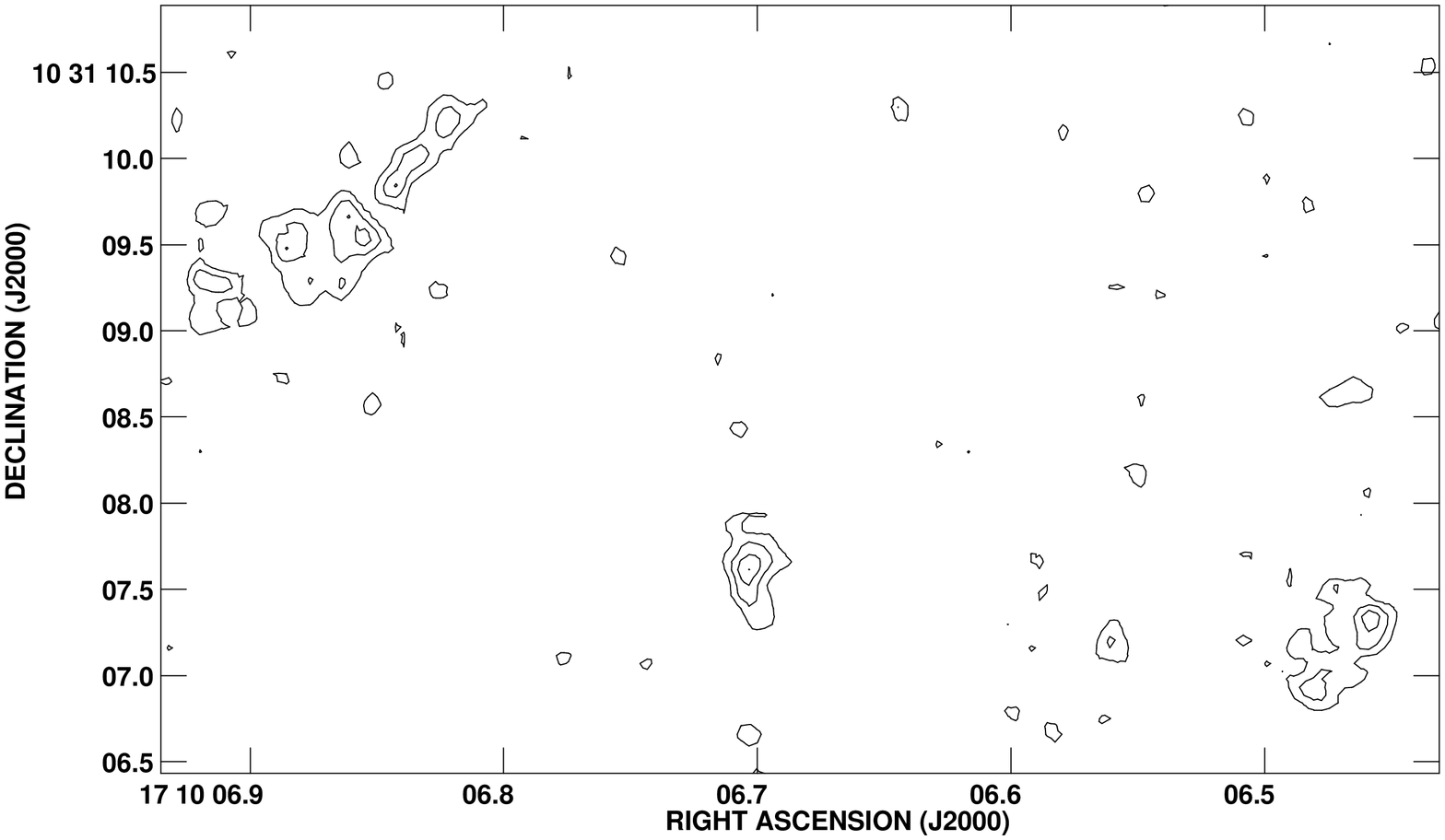,width=8.5cm,angle=-90}
}
\vskip-1cm
\centerline{
\psfig{file=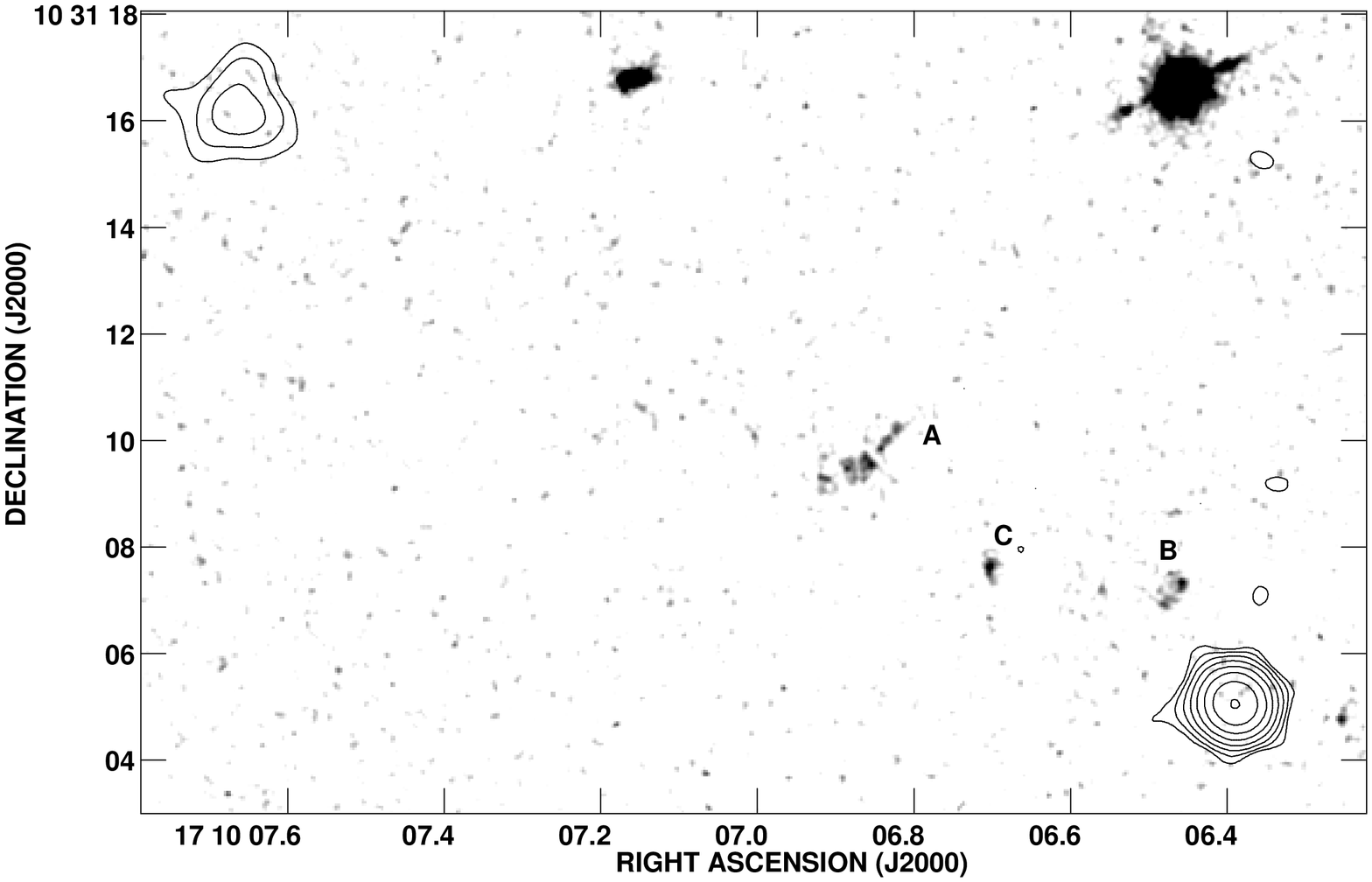,width=17.cm,angle=-90}
}
\vskip-1cm
\caption{ {\it Top}: 
 Grey scale image of TX 1707+105 at z=2.349.
{\it Center}: Contour representation of the UV continuum emission. Contour levels: 10,11,12,13,14$\times 6 10^{-22}$ erg s$^{-1}$ \AA$^{-1}$.
{\it Bottom}: Full image of the field of the radio source, with 
VLA radio contours superimposed.
Contours are at a geometrical sequence in 2 with the first contour at 
0.18 mJy. }
\end{figure*}
It  
 consists of a compact nucleus, a second bright component
and extended lower surface brightness emission which is clumpy. 
The galaxy and the radio source are strongly misaligned:
the angle between the optical and radio axis is nearly 45 degrees.
However, the northern component of the radio source is curved, suggesting the 
 radio axis might be precessing, in which case the elongated optical emission
could be located along the previous path of the radio jet.
The galaxy has extended ($\sim 80$ kpc) bright Ly$\alpha$ emission, 
exhibiting  a velocity shear that could be due to rotation of the gas.
The amplitude of this shear is almost equal to the overall velocity width of the line (van Ojik et al. 1997).  
\section{\bf HZRGs morphologies and evolution}
\subsection{\bf Radio optical alignment}
The UV-optical continuum emission from HZRGs is generally 
aligned with the main axis of the radio emission;
several models have been proposed to explain the nature of the optical continuum 
emission and of this alignment effect (for a review see  McCarthy 1993  
and references therein).\nocite{mcc93} 
The most viable ones are: (i) star-formation stimulated 
by the radio jets as it propagates outward from the nucleus 
(\cite{cha87,mcc87a,you89,dal90}); (ii) 
scattering of light from an obscured nucleus by dust or free 
electrons (di Serego Alighieri et al 1989; Scarrott et al 1990; Tadhunter et
al 1992; Cimatti et al 1993; di Serego Alighieri et al 1994); 
(iii) nebular continuum emission 
from warm line emitting clouds excited by the obscured nucleus (\cite{dic95}).
\\
\begin{figure}
\centerline{
\psfig{file=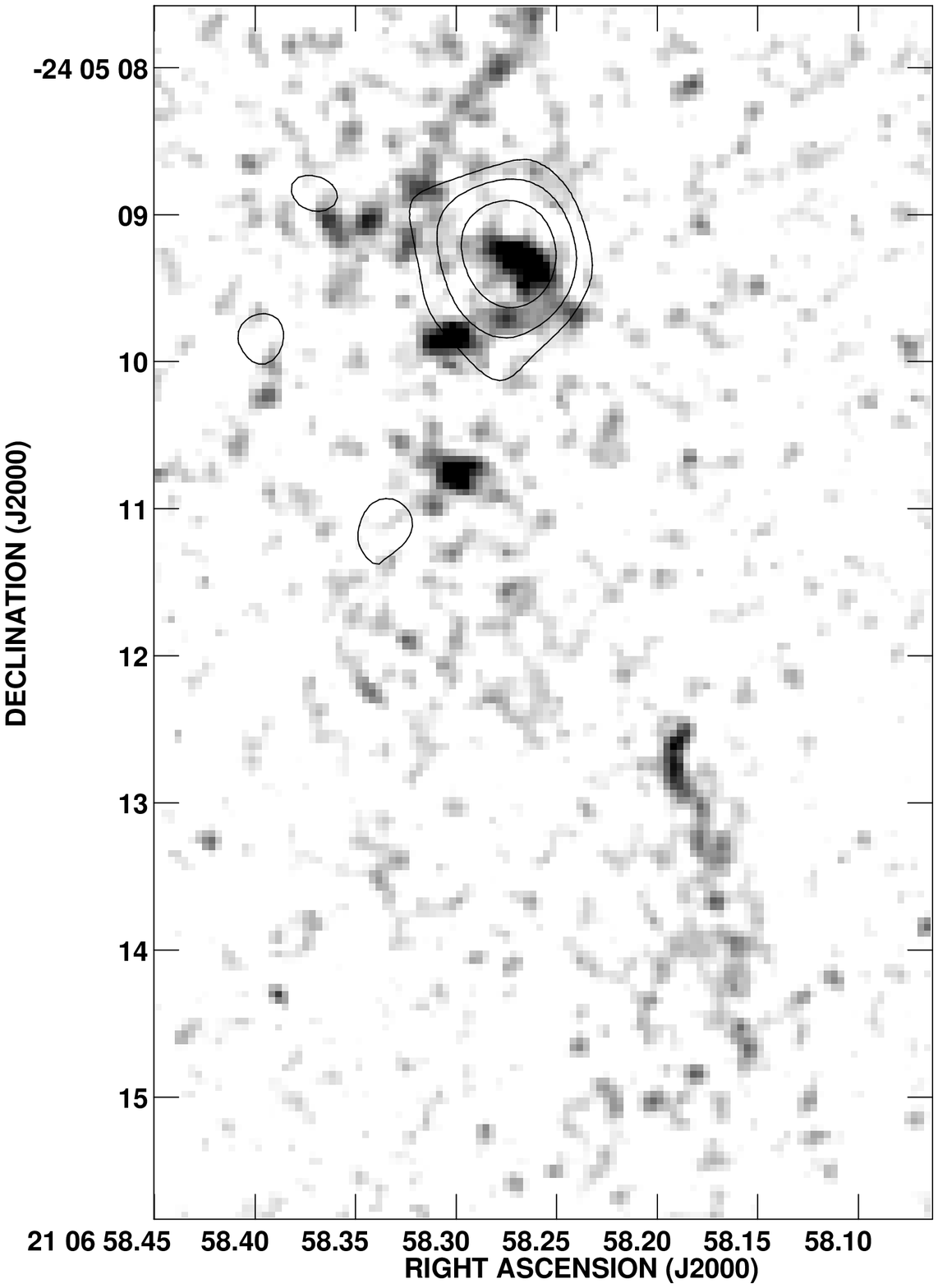,width=7cm}
}
\caption{ Grey scale image of
MRC 2104-242 at z=2.491, with VLA radio contours superimposed.
Contours are at a geometrical sequence in 2 with the first contour at 
0.07 mJy. }
\end{figure}
So far the only HZRG for which 
there is direct spectroscopic evidence that the UV continuum clumps are 
star forming regions, not dominated by scattered light, is 4C41.17: 
the spectrum of this galaxy shows absorption lines 
and P-Cygni profiles similar to those found in the spectra of high redshift 
star forming galaxies (Dey et al. 1997).
\\
Until recently, polarization measurements were possible only for z$\sim 1$
radio galaxies and showed that in most cases a large fraction of 
the UV continuum emission 
could be explained as scattered light. 
Recently though, observations of $z \ge 2$ radio galaxies have led to 
quite contradictory results: while some objects show considerable amounts of 
polarization (e.g Cimatti et al. 1997 and references therein), 
others  such as 4C41.17  have upper limits consistent with the complete absence of  polarization (\cite{dey97}). \nocite{cim97}
\\
The HST data with their high resolution 
provide informations about  the inner regions of HZRGs, and confirm that 
the radio/optical alignment is still present at scales of less than an  arcsecond.
In Fig. 12 we plot
the distribution of position angle difference ($\Delta \Theta$) for
our sample. To determine the optical position 
angle (PA), for the galaxies with a 
more regular morphology we smoothed the HST images with a Gaussian 
function having a FWHM of 1$''$ and then we fit the inner 3$''$ region 
with  ellipses, using the IRAF package  ISOPHOTE, which also gives the 
orientation  of the major axis of the ellipse. For the galaxies with 
irregular morphologies, 
the fits gave meaningless results,
so we selected as optical 
axis the line passing through the 2 brightest peaks on the images.
%
\begin{figure*}
\centerline{
\psfig{file=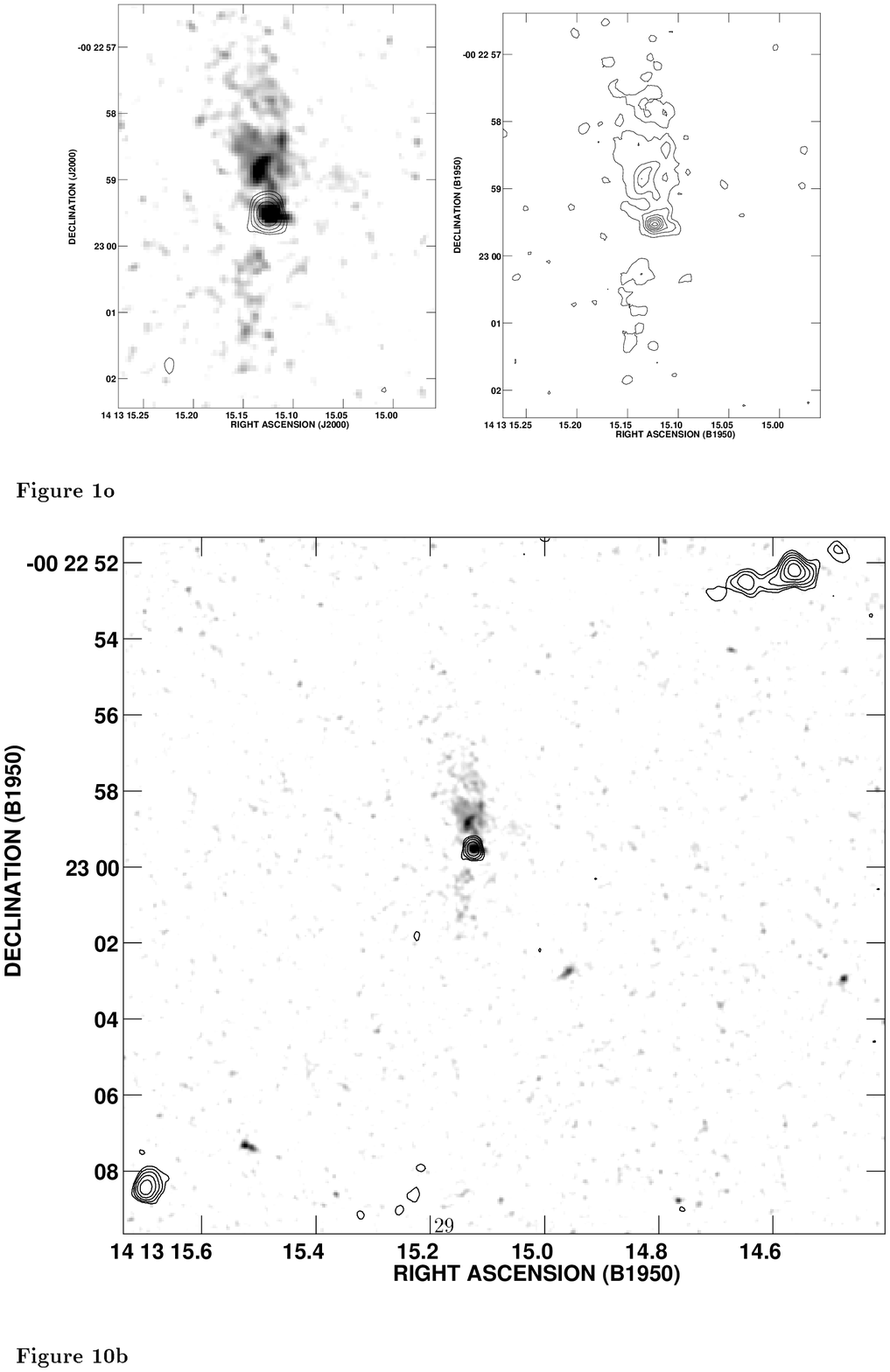,width=7.cm,clip=}
\psfig{file=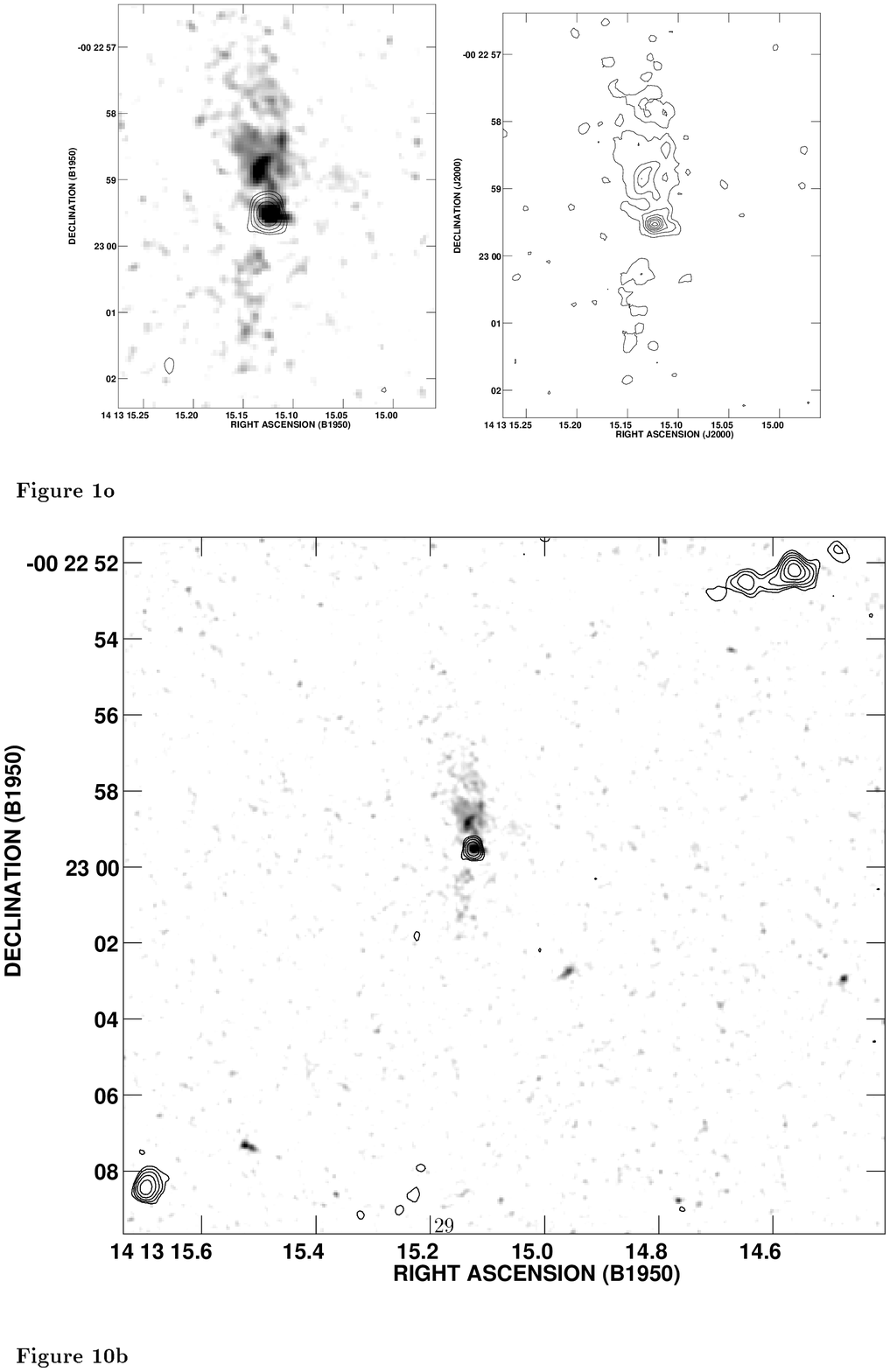,width=7.cm,clip=}
}
\vskip-1cm
\centerline{
\psfig{file=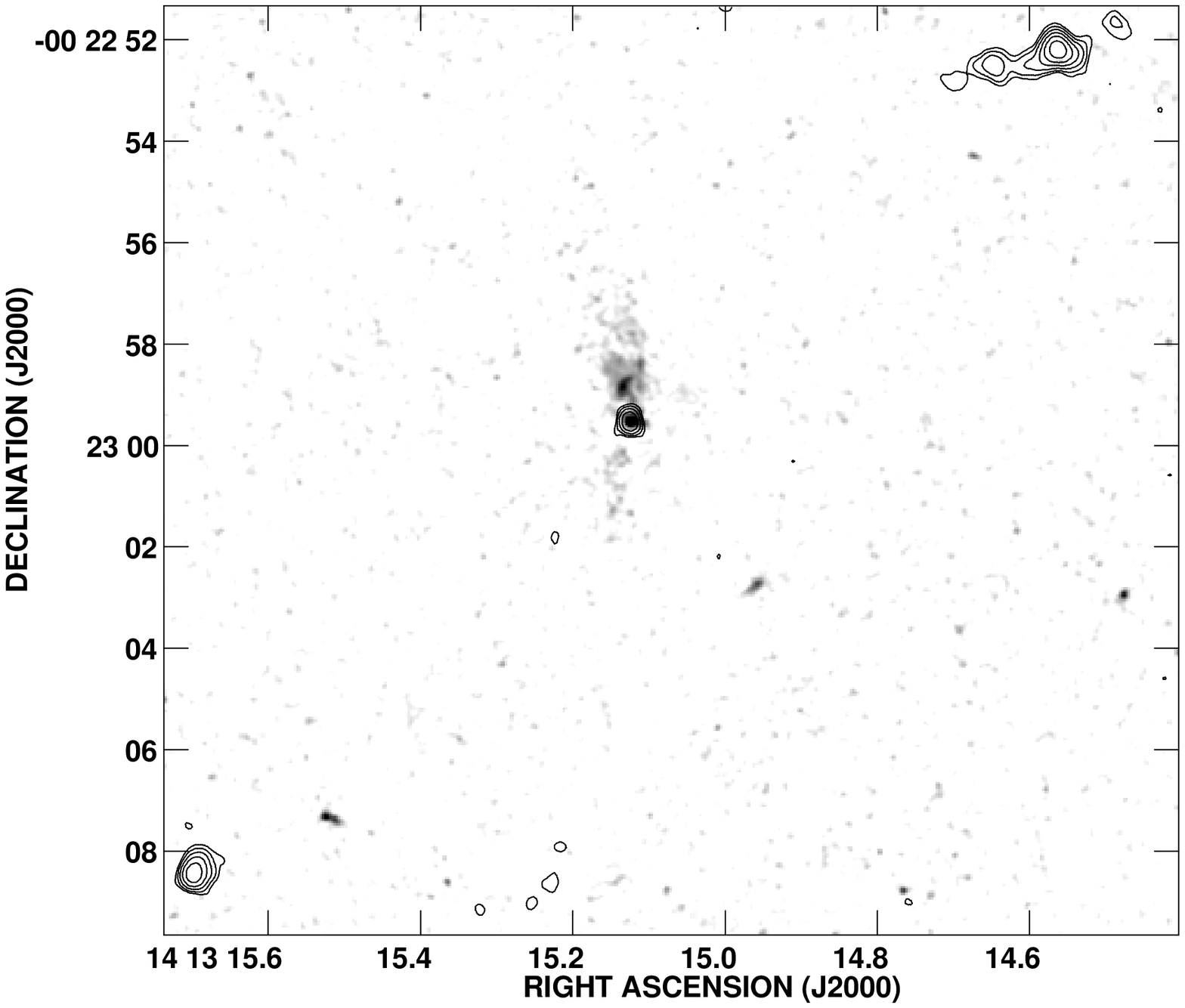,width=17.cm}
}
\vskip-1.5cm
\caption{{\it Top left}: Grey scale image of
4C 1410-001 at z=2.363, with VLA radio contours superimposed.
Contours are at a geometrical sequence in 2 with the first contour at 
0.11 mJy. {\it Top right}:  Contour representation of the UV continuum emission. Contour levels:10,11,12,13,14,15,16  $\times 1.2
 10^{-21}$ erg s$^{-1}$ \AA$^{-1}$. 
{\it Bottom}: An image of the full field of the radio galaxy. 
Radio contours and grey scale are the same as the upper left panel. }
\vskip-1cm
\end{figure*}
\begin{figure}
\centerline{
\psfig{file=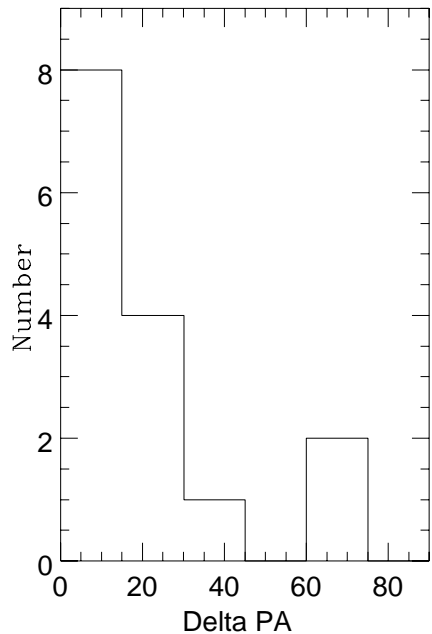,width=7cm,clip=}
}
\caption{Distribution of the differences in position angles 
between the radio emission and the UV continuum emission, measured in 
the inner 3$''$ region of the radio galaxies. The histogram includes data from the enlarged sample (see text).}
\end{figure} 
The position angle of the radio emission is given by the line joining the 
radio core to the nearest hot-spots (or the line joining the hotspots, 
if the core is not detected).
\\
Despite the fact that 13 out of 15 radio galaxies in our sample
have $\Delta \Theta \le 45^{\circ}$,  we notice that the properties 
of the alignment effect vary
considerably from object to objects. 
\\
We can distinguish various groups:
\\
(i) Radio galaxies that show a remarkable one-to one relation between 
radio emission and UV continuum light: 
this includes 
4C 1345+245, that has an optical jet-like feature, and 4C 1243+036 where 
the UV light follows the bending of  the radio jet.
These structures can be easily explained by the
jet-induced star formation models (see references above). 
Alternatively Bremer et al. (1997) proposed a mechanism by which,
when the radio jet passes through the gas clouds, it breaks them apart
thus increasing the surface area of cool gas exposed to the ionizing beam.
Consequently the material along the jet path becomes a far more efficient 
scatterer of nuclear radiation and the UV emission is enhanced in 
a  very narrow region. \nocite{bre97} 
\\
(ii) Radio galaxies in which the UV continuum emission
has a triangular-shaped  morphology, reminiscent of an ionization cone.
This category includes TX 0828+193, MRC 2025$-$218 and MRC 0406-242.
Such morphologies are expected in models 
that consider the aligned optical continuum  
as being scattered light of a central  buried quasar.
\\
(iii) Radio galaxies where the alignment between the optical morphology
and the radio axis is good but there no one-to  one relation between radio 
and UV components. This group includes MRC 0943$-$242, 
MRC 2104$-$242, 4C 1410$-$001, TX 0211$-$122, MG 2141+192, PKS 1138$-$262, 4C28.58 and 4C41.17. The degree to which the 
two components are aligned varies strongly even within this group: for example in  the radio galaxy 4C 1410-001 the difference in position angles between the radio and the UV emission is 45$^{\circ}$,
however the radio map indicates that the radio jets probably had a different 
direction in the past, corresponding to the direction along which the UV 
light is elongated.
\\
(iv) Finally, galaxies that show total misalignment 
between radio and optical emission. There are 2 such cases in our sample.
First the radio galaxy B2 0902+343 (Fig. 3), where
dust could play an important role in obscuring 
the central regions (Eales et al. 1993, Eisenhardt and Dickinson 1992),
thus ``masking'' the alignment effect. 
Second, the extremely peculiar and complex 
system TX 1707+105 (Fig. 9), which is comprised of 2 (possibly 3) 
separate galaxies, with similarly strong Ly$\alpha$ emission: the galaxies  
are located along the radio axis, but they are clumpy and extended 
almost perpendicularly to the radio axis. 
This unusual morphology would be hard to explain just by 
invoking the presence of dust, since the dust should have 
an extremely complicated distribution in multiple lanes 
parallel to the radio axis.    
\\
In summary the new data confirm that there is no single model
that can satisfactorily explain the optical morphology of all HZRGs
and the nature of the aligned optical continuum emission. 
At the same time, none of the proposed models can be ruled out by 
the present data.
Therefore it seems likely that all three mechanism 
contribute to the aligned light, 
but their relative importance  varies greatly from 
object to object.
\\
\subsection{\bf Clumpiness of the optical emission}
A striking feature of the HST 
images of the radio galaxies is the widespread clumpiness of the 
optical continuum emission. Most galaxies are comprised of several components, 
regardless of the fact whether they are aligned or not with the radio axis;
the clumps are resolved and their typical sizes are in the range 2-10 kpc.
\begin{figure}
\centerline{
\psfig{file=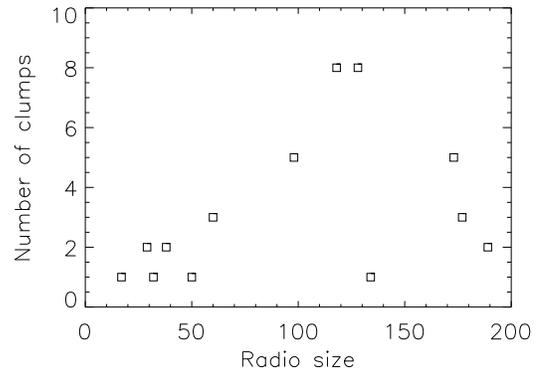,width=8cm}
}
\caption{Number of optical clumps of the galaxies versus total 
radio source size (in kpc).}
\end{figure} 
To give a consistent definition of ``clumpiness'' we proceeded in 
the following way: since the size of our  sample is small, and for the 
faintest galaxies it is difficult 
to delineate the structures,  
we first normalized the total observed flux of each galaxy 
(within a fixed aperture) 
taking the faintest and most distant  galaxy MG 2141+192
as reference.
We then  defined the parameter n  as the number of components which
have at least one contour at a flux level of
$4.4\times 10^{-19}(1+z)^{-4}$ erg cm$^{-2}$ sec$^{-1}$\AA$^{-1}$.
This value was chosen so that the radio galaxy MG 2141+192 had 3 clumps.
Note that, despite the difference in restframe 
frequencies sampled by the observations (see Table 1), this is a good approximation because  the 
spectral energy distribution of HZRGs in the UV wavelength range
(1300-2000 \AA) is generally flat.
\\
In Fig. 13  we present
a plot showing how  n, our measure of clumpiness, varies with  radio size,
for all the radio galaxies in the sample. Clearly there is a tendency for 
the larger radio sources
to have a clumpier optical continuum: 
the sources with radio sizes greater 
than $\sim 80$kpc have on average more than twice as many clumps as the 
smaller radio galaxies.
A Spearman rank correlation test gives a significance level of 95$\%$ for
this correlation.
\\
A possible explanation for this trend is that the 
medium around the hosts of powerful AGN is  dense and clumpy
on a scale of more than 100 kpc; as the radio sources expand 
through the gas, they light up  more and more material  either  
by triggering star formation 
in the gas clouds, or by enhancing the scattering properties 
of the material in the vicinity of the jets.
This result is
contrary to that found by Best et al. (1996) for
a complete sample of $z \simeq 1$  3CR radio sources, which have been imaged 
with the HST: they found that smaller radio sources tend to be comprised 
of a string of several knots, while larger radio galaxies 
are made generally of only 2 optical components.\nocite{bes96}
However note that the range of radio sizes of the  
$z \simeq 1$ 3CR  sample is 3 times as large as that of our sample.
\subsection{\bf Morphological evolution}
Our sample  covers a 
redshift range  from z$=$2 to z$=$3.8, which correspond to 
look-back times from 80\% to  90\% of the total age of the universe 
(for $\Omega =1$).  
This epoch is close to the epoch of the  formation of these HZRGS, 
therefore it is interesting to search for any evolution 
in the properties of the radio galaxies with increasing cosmic time.
We follow a similar approach to that used by  van Breugel et al. 
(1998) for a sample of powerful HZRGs, observed with Keck
in the near infrared, which corresponds to the
restframe optical emission ($>4000$ \AA). 
Their sample is similar to ours, being comprised or 
a similar number of sources, with the same radio power,  
but has a higher average redshift  ($z_{av}= 3.2$ versus $z_{av}=2.8$) 
and more galaxies having $z \ge 3$. There are 6 radio galaxies 
common to both samples.
\\
In Fig. 14 we present the results for our sample of HZRGs: the left plot 
shows how the radio /optical size ratio 
vary with redshift; the radio sizes are measured as the 
distances between the most distant 
hot-spots, on either side of the nucleus, while the optical lengths 
are defined to be the maximum extension of the optical emission 
in the direction of the radio source.  
In cases of multiple systems, such as PKS 1138-262 and TX 1707+105, all the 
optical components where considered, so that the  radio/optical size 
ratio gives an indication of how much emission there is within the 
radio source extension.
The plot indicates that there is  no significant evolution in the ratio
 of radio to optical  size. 
If we divide the sample in two redshift ranges, 
then the average radio/optical size ratio is 3.2 for the highest 
redshift bin ($z \ge 2.9$ ) and 3.4 for the lowest redshift 
radio galaxies, so the difference is negligible.
This is different from the result of an Breugel et al. for the 
infrared emission:  they present marginal evidence  that the host of $z \ge 3$ radio galaxies 
are comparable in size with the radio sources, while the $z \le 3$  
radio sources appear systematically larger that the hosts galaxies.
\\
In the right plot of Fig. 14  we show how the strength of the radio-optical 
alignment, represented by  the difference in position angle between the
optical and the radio emission $\Delta \Theta$ 
(see Sect. above for definition) varies with z. 
Again there is no significant difference between the lowest redshift radio 
galaxies,  which  have an average PA difference of 20$^{\circ} 
\pm 7^{\circ}$\footnote{We preferred not to include 
the radio source TX 1707+105 in calculating the average PA of the low redshift group, because for this source 
the PA of the single  galaxy 1707+105A, ($\sim69^{\circ}$), is extremely different from the PA  of whole system (3 galaxies, PA$\sim13^{\circ}$)}
and the highest redshift sources 
which have an average of 18$^{\circ} \pm  8^{\circ}$.
On the contrary van Breugel et al. find a strong evolution in 
the alignment of the host galaxies from $z \ge 3$ to $ z \le 3$: 
specifically the infrared morphologies become smoother and 
less elongated at  $ z \le 3$ and the infrared/radio alignment strength decreases. The best interpretation is that, while for the lower 
redshift sources in their sample the near IR emission is dominated by the most 
evolved stellar population, (which is less effected by the presence of the 
radio jets), for the very high redshift galaxies 
the observed near IR emission starts to be dominated by young stars,
probably formed following the passage of the radio jets.
\\
On the other hand, our HST observations sample 
the UV restframe emission which is though to be dominated by
the younger stellar populations 
in all cases, regardless of redshift.
These young hot stars are formed in subsequent small bursts,
induced either by the interaction of the jets with the medium 
or by mergers of smaller subunits. Such events may 
involve only little amounts of mass, but can still produce remarkable 
UV morphologies (e.g 1138-262 Pentericci et al. 1998), and 
their frequency is not expected to change from redshift 4 to 2.
Therefore we don't expect any strong evolution in the UV restframe 
properties of the radio galaxies in this redshift interval.
\\
\\
\begin{figure}
\mbox{
\psfig{file=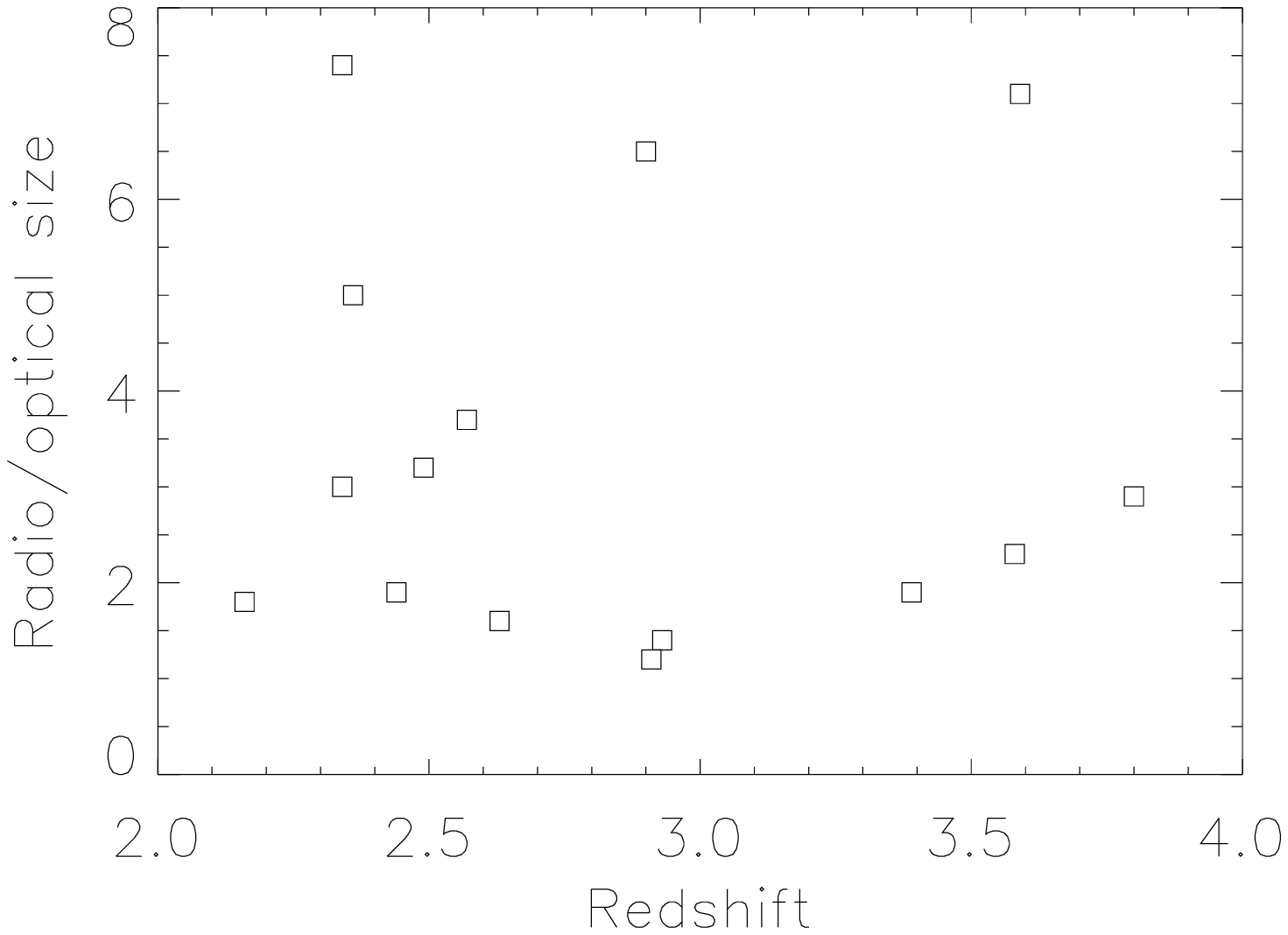,width=7cm,clip=}}
\mbox{
\psfig{file=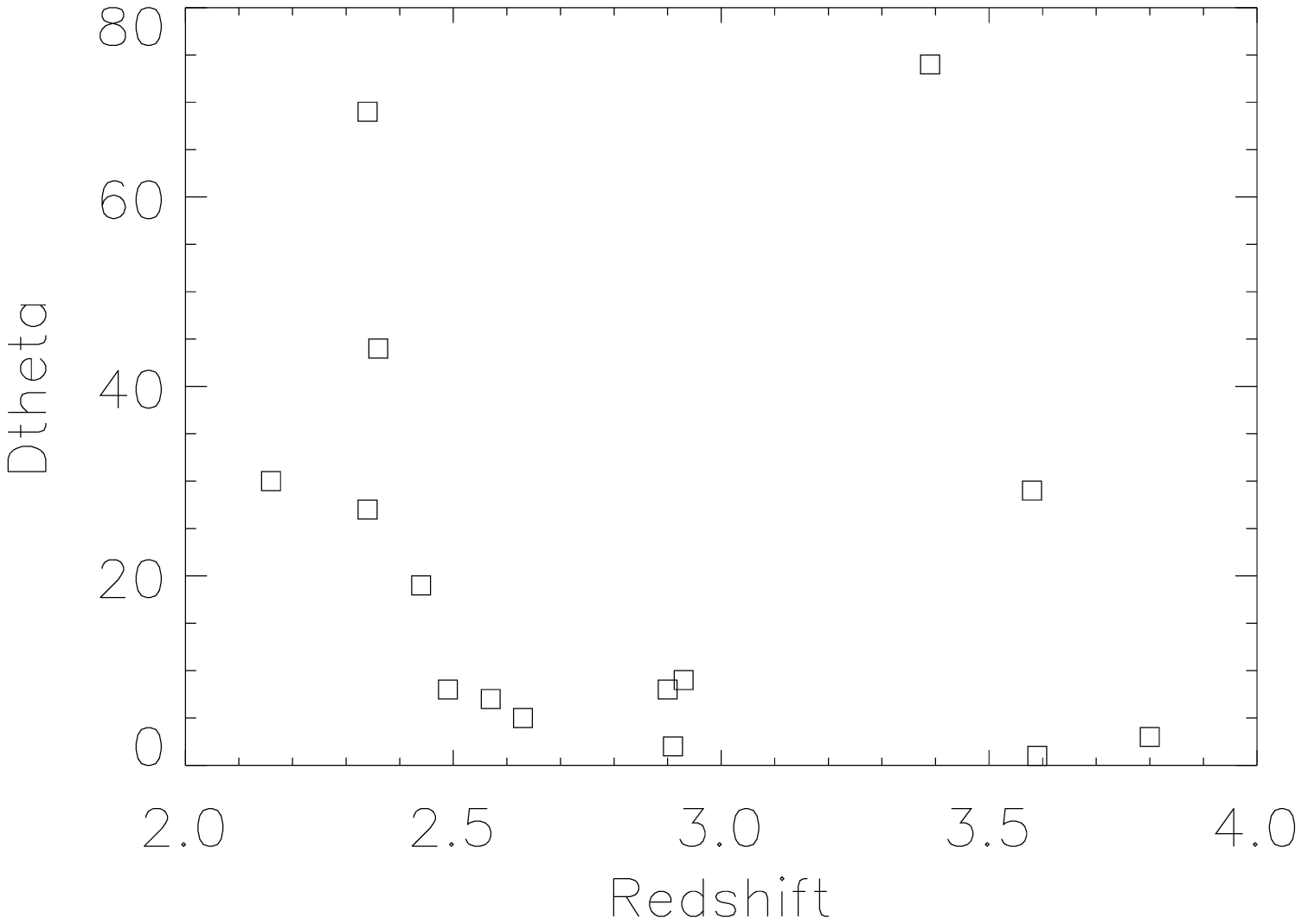,width=7cm,clip=}}
\caption{Evolution of the alignment strength (upper panel) 
and the radio/optical  size ratio (lower panel) with redshift.}
\end{figure} 
\subsection{\bf The formation of brightest cluster galaxies?} 
It is interesting to compare the morphologies of our 
high redshift radio galaxies with those of the high redshift galaxies,
which have been recently discovered with  UV dropout techniques
and  extensively studied by various groups, also with the HST
(see for example Steidel et al. 1996, Giavalisco et al. 1996, 
Williams et al. 1996). 
\nocite{ste96,gia96} 
\\
As pointed out in Sect. 6.1, a fraction of the UV continuum emission
of HZRGs  is directly connected, through various possible mechanisms, 
to  the presence of the AGN.
Also some of the features that we see in the galaxies can be easily 
explained by a direct correlation with the radio jets, for example the 
narrow elongated structures seen in 4C 1243+036 and MRC 2104$-$242 and the 
jet-like feature observed in 4C 1345+245.
However in other cases there is a striking similarity between the 
{\it individual components}  of the radio galaxies  and 
the population of UV dropout galaxies,
which clearly favors a stellar origin for the emission coming 
from those clumps.
\\
Particularly in some of the clumpiest and most extended radio galaxies,
such as TX 1707+105 and MRC 2104$-$242, there are components that  
have a compact and regular morphology, with sizes of the order of few kpc
resembling that of the high redshift radio-quiet galaxies. 
In a previous paper we made a detailed inter-comparison between the clumps
that are observed around the radio galaxy 1138$-$262  
to the UV  dropout galaxies: the conclusion was that 
those  components had characteristics similar to the 
UV dropout galaxies  such as
absolute magnitudes, surface brightness profiles,
half-light radii
($\sim 2$ kpc) and inferred star  formation rates (5-10 $M_{\odot}$yr$^{-1}$ 
per clump; Pentericci et al. 1998).
Also  4C41.17 has a similar very clumpy morphology 
with compelling  evidence that the clumps are star forming regions. 
However we must note that it is not yet possible to determine 
the masses of neither of the 2 classes of objects (UV dropouts and radio galaxies clumps);
so it could as well be that they are intrinsically different objects, 
with a similar amount of star-bursting activity that make them look similar in the UV continuum emission.
\\
It seems that at least some of the high redshift 
radio galaxies consist of a central large
 galaxy, that hosts the AGN 
and a number of small star forming subunits, resembling  
the UV dropout galaxies, which are located in a region as large as 
$\sim 50-100$ kpc around the radio source. 
Powerful radio sources would then pin-point to regions
in which the density of star forming units is higher than average.
The central host galaxies of radio sources may well have formed  
trough merging of these small sub-galactic stellar systems. 
Note that the mergers of these gas-rich subunits 
with the host galaxies could have triggered (or re-triggered) 
the radio emission 
by providing fuel for the central engine of the AGNs, as it seen 
in many cases at low z (e.g Osterbrock 1993).
\\
Our observations provide some qualitative support for hierarchical 
galaxy evolution models, which predict that 
the morphological appearance of galaxies during their formation period 
should be highly irregular and clumpy (e.g Baron \& White 1989).
In particular  semi-analytical models predict that 
one of the forms in which massive elliptical galaxies
accrete their mass is from multiple merging of smaller subunits 
(Aragon-Salamanca et al. 1998, and 
references therein.
A possible problem arises from the fact that 
in  standard hierarchical cold dark matter models 
such massive systems are thought to form relatively late (\cite{col94,kau93}),
i.e. at much lower redshift, and in the majority of galaxies 
the main population of stars is formed more recently (after $z=1$) 
\cite{hey95}. 
However, White \& Frenk (1991) argue
 that a  mechanism that could explain the formation of massive 
elliptical  galaxies at an earlier epoch is over-merging  
of star-burst galaxies and indeed, as we have reviewed in the introduction,
there is now increasing evidence that 
high redshift radio galaxies are probably located in the over-dense 
regions of the early universe. 
\\
Therefore we conclude that high redshift radio galaxies 
may be  formed from a aggregates of sub-galactic units, similar to the 
UV dropout galaxies, 
and will probably evolve into present day  
brightest cluster galaxies.
\section{Summary and concluding remarks}
In this paper we have presented new HST/WFPC2 images of 
11 high redshift radio galaxies,
all complemented with VLA radio maps of comparable resolution.
The images reveal a wide variety in the morphology of the host galaxies
of these high redshift radio sources: in particular 
most objects have a clumpy, irregular appearance, consisting 
of a bright nucleus  and a number of smaller components. The number of clumps
seems to increase with increasing radio size.
The UV continuum emission is generally elongated and aligned with 
the axis of the radio sources, however the characteristics of the 
``alignment effect'' differ greatly from case to case. 
The new data confirm that none of the proposed models 
can satisfactorily explain the phenomenon and that most probably the 
aligned continuum emission is a mixture of star light, scattered light, and nebular continuum emission.
Our data show no significant evolution in the morphological properties 
over the redshift interval. Finally, we compare the properties of  
our radio galaxies with those of the 
 UV dropout galaxies and conclude that 
high redshift radio galaxies might be forming  from aggregates of sub-clumps
similar to the  UV dropout galaxies and that they will 
probably evolve into present day brightest  cluster galaxies. 
\\
In a future paper we will present complementary HST/NICMOS data 
of an enlarged sample of high redshift radio galaxies. 
The new infrared observations will provide constrains to the age of the 
older stellar population of the host galaxies. 
With the high resolution we will be able to determine 
if also the older stellar population shows significant 
clumpy sub-structures and to  what extent are the forming 
brightest cluster ellipticals already assembled and relaxed.
\begin{acknowledgements}
This work is based on observations with the NASA/ESA Hubble Space Telescope,
obtained at the Space Telescope Science Institute, which is operated by AURA
Inc. under contract with NASA. 
We thank C. Carilli for doing the reduction of the radio data for TX 1707+105.
HJAR acknowledges support from an EU
twinning project, a programme subsidy granted by the Netherlands
Organization for Scientific Research (NWO) and a NATO research grant.
The work by WvB at IGPP/LLNL was performed under the auspices of the US Department of Energy under contract W-7405-ENG-48.
\end{acknowledgements}
\newpage
\begin{table*}
\caption{\large{\bf Radio properties of TX 1707+105}}\label{tabap.pol}
\hskip1cm
\begin{tabular}{lcccccccccc}
\hline
Component& S$_{8.4}$&S$_{4.8}$&I$_{8.4}$&I$_{4.8}$&$\alpha$&S$^{p}_{8.4}$&
S$^{p}_{4.8}$ & FP$_{8.4}$ & FP$_{4.8}$ & RM
\\ 
         & mJy      & mJy     &mJy/beam &mJy/beam &        & mJy& mJy& \% &\% &
rad m$^{-2}$
\\
\\
NW       & 2.85     & 7.47    &1.07     &4.69     & 2.6    &    &    &   & 
\\  
SE       & 27.3     & 55.8    &23.4     &53.1     & 1.5    &4.85&4.12& 18&7.4
& 47
\\
\hline
\end{tabular}
\end{table*}
\begin{table*}
\caption{\large{\bf Radio properties of MRC 2104-242}}\label{tabap.pol}
\begin{tabular}{lcccccccc}
\hline
Component  &S$_{8.2}$&S$_{4.8}$&S$_{1.5}$&I$_{8.2}$&I$_{4.8}$&I$_{1.5}$&
$\alpha_l$&  $\alpha_h$   
\\
           & mJy     &mJy      &  mJy    &mJy/beam &mJy/beam &mJy/beam &
\\
\\
North lobe & 22.9    &55.2     & 294     & 10.4    & 49.8    & 278     &1.45&2.9
\\   
Core       & 0.55    &         &         & 0.53    &         &         & 
\\
South lobe & 1.77    &7.14     & 34.8    & 1.27    & 4.66    & 19.2    & 1.2& 2.5
\\
\hline
\end{tabular}
\end{table*}
\begin{appendix}
\section{\bf Radio images of TX 1707+105}
We present here multi-frequency maps of the radio
galaxy TX 1707+105 obtained with the VLA in B array.
\\
Observations were made at 4.5 and 8.2 GHz, using two frequency channel each
having a  50 MHz bandwidth, 
 for a total integration time of 700 s and 1020 s respectively.    
 Data processing was done performed using the 
Astronomical Image Processing System (AIPS) in the standard way. 
The system gains
were calibrated with respect to the standard sources 3C286. 
Phase calibration was performed using the nearby calibrator 1658+076. 
The antenna polarization response terms were determined using multiple scans of the calibrator 1850+284 over a large range in parallactic angle.
 Absolute linear polarization position angles were measured using  
a scan of  3C286. The calibrated data were then edited and self-calibrated
using standard procedures to improve the dynamic range. 
Images of the three Stokes polarization parameters, I, Q and U  were 
synthesized and all images were CLEANed down to a level of approximately
3 times the theoretical rms noise using the AIPS task IMAGR. The observations
at the different frequencies were added in the image plane to produce the 
the final maps of total and polarized flux.
\\
In Fig. 1 we show the maps at 4.5 GHz and 8.2 GHz (with a resolution 
respectively of 1.2$''$ and 0.7$''$) 
of the total flux (left panels) and polarized flux (right panels).
In all images contours are spaced in a geometric progression with a 
factor of 2, with the first contour level equal to 3$\sigma$, 
where $\sigma$ the off-source  rms which is  0.12 mJy for the 4.5 GHz map, 0.1 mJy for the 8.2 GHz map and 0.17 mJy for the polarized 
flux maps.
\\
The radio galaxy has a simple double morphology with no radio core detected
in the present images.
The two lobes are nearly symmetric in total radio brightness, but
the northern hot-spot is  totally depolarized at both frequencies, 
while the southern one is polarized.
\\
\section{\bf Radio images of MRC 2104-242}
In Fig. 2 we present maps of the radio galaxy 
MRC 2104$-$242 obtained with the VLA in B array at 3 different frequencies: 
1.4 GHz, 4.5 GHz, and 8.2 GHz, with a resolution, respectively of 3.9$''$, 
1.2$''$ and 0.7$''$.
In all images contours are spaced in a geometric progression with a 
factor of 2, with the first contour lever equal to 3$\sigma$, 
where $\sigma$ the off-source  rms, and is respectively at 1.74 mJy for the 1.5 GHz map, 0.19 mJy for the 4.5GHz map and 0.05 mJy  for the 8.2 GHz map.
\\
The radio source is a double showing  fainter diffuse emission between the hot-spots and the core. The northern hot-spot is elongated in a direction the is 
different from the radio axis.
\begin{figure*}
\centerline{
\psfig{figure=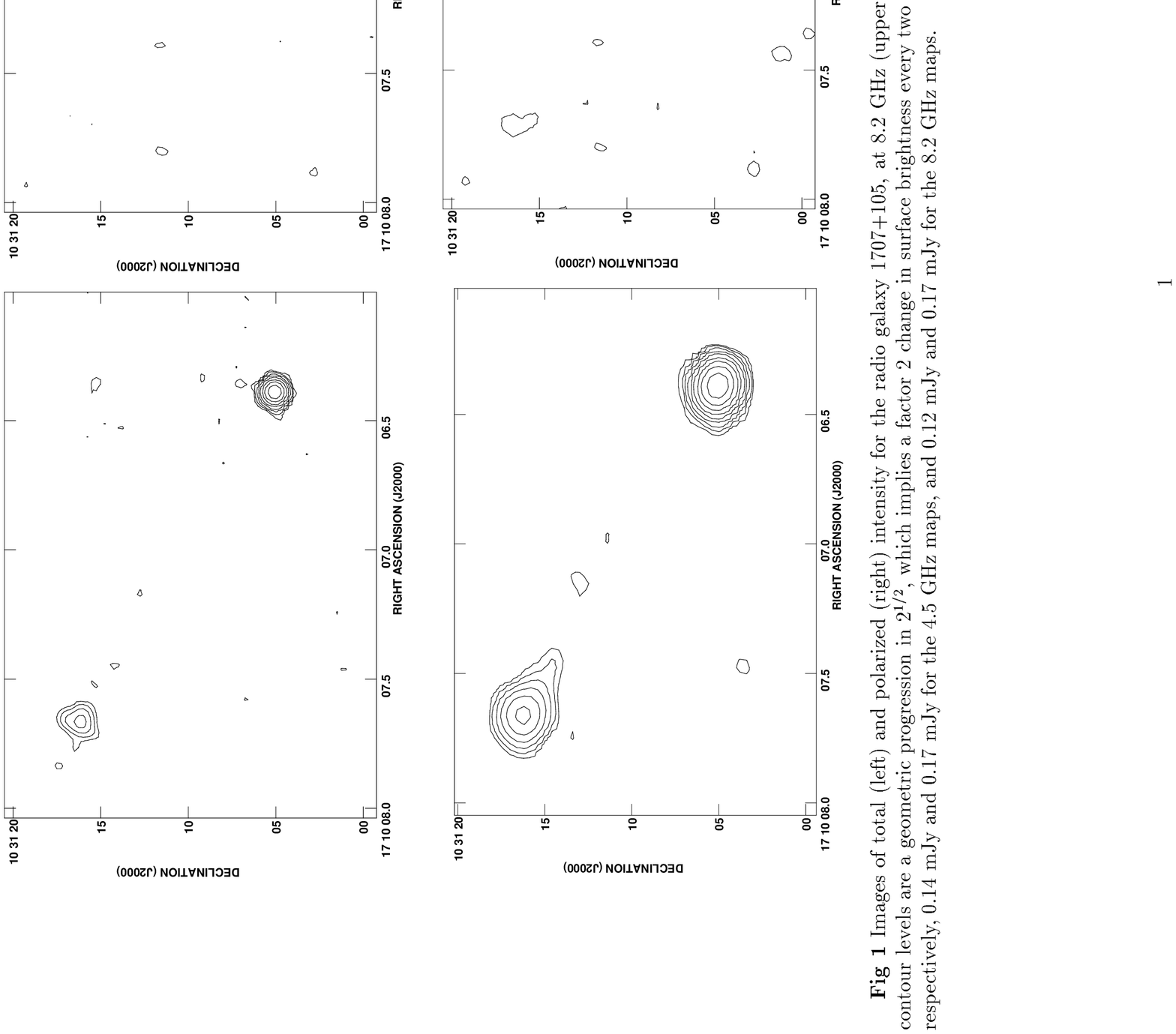,width=15.4cm,angle=90,clip=}
}
\end{figure*}
\begin{figure*}
\centerline{
\psfig{figure=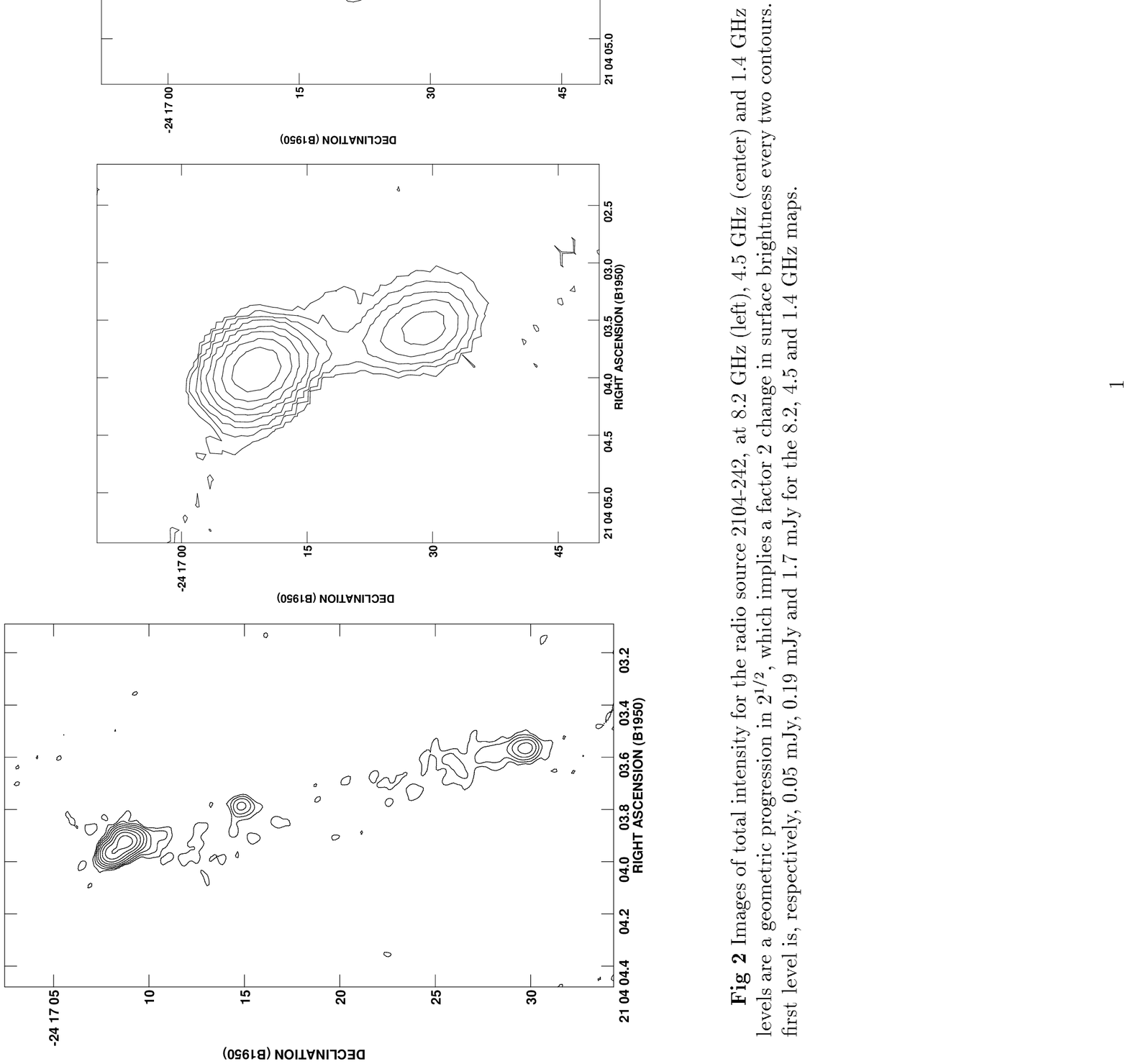,width=20cm,angle=90,clip=}
}
\end{figure*}
\end{appendix}

\end{document}